\documentclass[11pt]{article}
\usepackage{amsmath, amsthm}
\usepackage{amsmath}
\usepackage{amsfonts}
\usepackage{amssymb}
\usepackage{mathrsfs}
\usepackage{verbatim}
\usepackage[pdftex]{graphicx}
\usepackage{wrapfig}
\usepackage{amscd}
\usepackage{amsthm}
\usepackage{color}

\newtheoremstyle{thm}{1.5ex}{1.5ex}{\itshape\rmfamily}{}
{\bfseries\rmfamily}{}{2ex}{}

\newtheoremstyle{rem}{1.3ex}{1.3ex}{\rmfamily}{} 
{\itshape}
{} {1.5ex}{}

\newtheorem{thm}{Theorem}[section]

\newtheorem{cor}[thm]{Corollary}
\newtheorem{lemma}[thm]{Lemma}

\theoremstyle{definition}
\newtheorem{remark}[thm] {Remark}
\newtheorem{nota}[thm]{Notation}
\newtheorem{defn}[thm]{Definition}
\newtheorem{ex}[thm]{Example}

\numberwithin{equation}{section}

\DeclareFontFamily{OT1}{pzc}{}
\DeclareFontShape{OT1}{pzc}{m}{it}
             {<-> s * [0.900] pzcmi7t}{}
\DeclareMathAlphabet{\mathscrb}{OT1}{pzc}
                                 {m}{it}
                                 
\definecolor{purple}{rgb}{1, 0, 1}

\begin{document}

\title{\Large{\textsc{Cardy's Formula for Certain Models of the Bond--Triangular Type}}}

\author
{\large{L.~Chayes$^1$ and H.~K.~Lei$^1$}\thanks{\copyright\, 2006 by L.~Chayes and H.~K.~Lei.  Reproduction, by any means, of the entire article for non-commercial purposes is permitted without charge.}}

\maketitle

\vspace{-4mm}
\centerline{${}^1$\textit{Department of Mathematics, UCLA}}

\begin{quote}
{\footnotesize {\bf Abstract:} We introduce and study a family of 2D percolation systems which are based on the bond percolation model of the triangular lattice.  The system under study has local correlations, however, bonds separated by a few lattice spacings act independently of one another.  By avoiding explicit use of microscopic paths, it is first established that the model possesses the typical attributes which are indicative of critical behavior in 2D percolation problems.  Subsequently, the so called Cardy--Carleson functions are demonstrated to satisfy, in the continuum limit, Cardy's formula for crossing probabilities.  This extends the results of S.~Smirnov to a non--trivial class of critical 2D percolation systems.}   

{\footnotesize {\bf Keywords:} Universality, Conformal invariance, Cardy's formula, Critical percolation.}
\end{quote}

\section{\large{Introduction}}
\subsection{\large{Introductory Remarks}}

In recent years, tremendous progress has been made towards understanding the (limiting) behavior of critical 2D percolation models; much of this is contained in the works of \cite{Smir}, \cite{NewC}, \cite{SmirW}, \cite{LSW}.  However, with very few exceptions, e.g.~long distance behavior of certain multi--arm correlations \cite{LSW}, \cite {AizChi}, \cite{KSZ}, all results have been confined to the site percolation model on the triangular lattice and scaling limits thereof.  Indeed, as uncovered by Smirnov \cite{Smir}, on this particular lattice, there is a miraculous local $120^\circ$ symmetry that facilitates the passage to the continuum.  Needless to say, an underlying theme behind ``invariant critical behavior'' is some notion of \emph{universal} behavior for the limiting model.  Unfortunately, the problem of extending Smirnov's result to other well--known 2D percolation models has, so far, proved illusive.  Here we present some limited progress towards these goals by establishing that in addition to the site problem on the triangular lattice, Cardy's formula holds for a modified bond problem on the triangular lattice.

We remark that in \cite{NewC1} and \cite{NewC2}, some steps in this direction have already been taken.  However, the critical models considered therein were, at long distance, demonstrably equivalent to the triangular site model from which they were evolved.  In particular, the asymptotic behavior of the connectivity functions and the cluster size distributions can be bounded above and below by their counterparts from the independent model on the triangular site lattice.  Thus the mere existence of ``$\eta$'' and ``$\delta$'' for the independent site model (implied by \cite{Smir}, \cite{NewC}, \cite{SmirW}, \cite{LSW}) gives this for free in the models of \cite{NewC1} and \cite{NewC2}.  This deviates somewhat from the original spirit of scaling and universality: it is supposed that one can \emph{infer} the critical exponents of a given lattice model via the universality class to which it belongs.  

The work of the present note is in rather closer adherence to the above--mentioned order of reasoning.  We construct a model based more on triangular bond percolation than site percolation.   (For technical as well as aesthetic reasons, local correlations between neighboring bonds will be introduced, but all events separated by three or more lattice spacings are independent.)  While perhaps obvious on the level of heuristics, critical behavior of the model requires verification; indeed this constitutes a non--trivial fraction of the work.  When this is achieved -- around the end of Section 2 -- one has a fairly standard--looking percolation--like model, not particularly distinguished from the myriad of critical 2D percolation models which one presumes is equivalent, in the scaling limit, to the limit obtained from the site model on the triangular lattice.  We remark, however, that before the advent of this work, and as likely as not in its aftermath, this will be among the less well--known models of critical 2D percolation.  Notwithstanding a derivation for this model, which parallels the derivation in \cite{Smir}, is obtained for universal -- and conformally invariant -- behavior of the limiting crossing probabilities.       

\subsection{\large{Background and Smirnov's Proof}}
In \cite{Smir}, a conformal invariant was found for critical site percolation on the triangular lattice that amounts to the conformal invariance of certain crossing probabilities and a verification of Cardy's formula \cite{Cardy}.  These properties allow the unique determination of the scaling limit \cite{Werner} via a connection to \textsc{SLE}$_6$.  As our general strategy follows closely that of \cite{Smir}, we include here a short discussion on \cite{Smir} and set up some general notation -- before launching into the specifics of our problem in the next section.  We will be succinct since most of what we say here can be found in the first part of \cite{Smir}.

Let $\Lambda$ denote a piecewise smooth domain which is the conformal image of a triangle.  We denote the portions of the boundaries corresponding to the sides of the triangle by $\mathscr A$, $\mathscr B$ and $\mathscr C$, and the associated vertices by $e_{_{AB}}$, $e_{_{BC}}$ and $e_{_{CA}}$ respectively.  The sequence ($\mathscr A, e_{_{AB}}, \mathscr B, e_{_{BC}}, \mathscr C, e_{_{CA}}$) should be regarded as counterclockwise ordered.  

Let $h_{_A}$, $h_{_B}$ and $h_{_C}$ denote the linear and hence harmonic functions defined on the unit equilateral triangle with vertices at $z=0$, $z=1$ and $z=\frac{1}{2} + i\frac{\sqrt{3}}{2}$: 
\begin{equation*}
h_{_A} = 1 - (x + \frac{1}{\sqrt{3}} y), \hspace{3mm} h_{_B} = x-\frac{1}{\sqrt{3}}y, \hspace{3mm} h_{_C} = \frac{2}{\sqrt{3}}y.
\end{equation*}  
Notice that $h_{_A}$ vanishes on one of the boundaries (the $\mathscr A$ boundary) and is equal to one at the vertex $e_{_{BC}}$, and similarly for $h_{_B}$ and $h_{_C}$.  Let $h_{\mathscr A}$, $h_{\mathscr B}$ and $h_{\mathscr C}$ denote the corresponding functions under the appropriate conformal transformation which takes the above--mentioned triangle into $\Lambda$.  Note that the boundary conditions, including the vertices are preserved under this transformation.  Obviously, even after the transformation, these three functions are not independent, e.g.~they add to one.  More importantly, they form a ``harmonic triple''; i.e. the functions 
\begin{equation*}
h_{\mathscr A} + \frac{i}{\sqrt{3}}(h_{\mathscr B} - h_{\mathscr C}), \hspace{3mm} h_{\mathscr B} + \frac{i}{\sqrt{3}}(h_{\mathscr C} - h_{\mathscr A}), \hspace{3mm} h_{\mathscr C} + \frac{i}{\sqrt{3}}(h_{\mathscr A} - h_{\mathscr B})
\end{equation*}
are all analytic.

\begin{defn}\label{functions}
Let $\Lambda$ and $\mathscr A$, etc.~be as above and consider the intersection of $\Lambda$ with the triangular site lattice with spacing $N^{-1}$.  Let us consider critical percolation on this lattice -- sites are blue or yellow with probability $\frac{1}{2}$ and, for $z \in \Lambda$, define $\mathscrb U_{_N}(z)$ to be the event that there is a path from $\mathscr A$ to $\mathscr B$ which separates $z$ from $\mathscr C$.  Similarly we define $\mathscrb V_{_N}$ and $\mathscrb W_{_N}$ cyclically.  We note that for each of the $u$, $v$ and $w$ there are in fact two objects to consider, namely a blue version of the event and a yellow version, but we will not let these details detract us from this informal discussion; similarly one should also define, with a bit of precision, the definition of the boundaries $\mathscr A$, $\mathscr B$ and $\mathscr C$ in accord with the lattice--approximation of $\Lambda$).  We let $u_{_N}$, $v_{_N}$ and $w_{_N}$ be the probabilities of the events $\mathscrb U_{_N}$, $\mathscrb V_{_N}$ and $\mathscrb W_{_N}$, respectively and consider the limits of these functions as $N \rightarrow \infty$ (if the limit indeed exists).    
\end{defn}  

The seminal result of the work by Smirnov \cite{Smir} is that as $N \rightarrow \infty$, each of these functions converge to the appropriate $h_\mathscr A$, $h_\mathscr B$ or $h_\mathscr C$ mentioned above.  We note that on the equilateral triangle these $h$'s (by definition) satisfy the Cardy--Carleson Formula and therefore they satisfy Cardy's formula on any conformal domain.  

Next we say a few words about the strategy for the proof of this theorem.  The lattice functions, which satisfy the same boundary conditions as the continuum $h$'s, are shown to converge, at least subsequentially.  Appropriate combinations of the limiting functions are demonstrated to be analytic, the key ingredient being a verification of the Cauchy condition for a (relatively) arbitrary contour.  Boundary conditions and some uniqueness arguments completely specify the limiting functions.   

The crucial ingredient which underpins the entire scheme is the existence of a set of Cauchy--Riemann type equations -- referred to as \emph{Cauchy--Riemann relations} -- which equate various directional derivatives of $u_{_N}$, $v_{_N}$ and $w_{_N}$ at the \emph{discrete} level.  In particular, the difference between any one of these functions at neighboring lattice sites has a probabilistic interpretation or, more precisely, may be expressed as the difference of two probabilities.  Both the positive and negative pieces of these derivatives are shown to be exactly equal to nearby counterparts of an appropriate member of the triple of functions.  Roughly speaking, (and here we refer the reader to the original reference \cite{Smir} or to Section 3 of the present note), the keynote of the strategy is ``color switching''.  Indeed, the derivative pieces turn out to be the probability of three paths emanating from the three boundaries and converging at the point where the derivative is taken.  The colors of the paths determine which particular function the derivative piece should be associated with.  Hence changing a path color changes the function and this amounts to a Cauchy--Riemann relation.  The ability to freely switch the colors of paths -- which is not common among the standard critical percolation models -- is an inherent symmetry of the triangular site percolation model at criticality.  

The major technical obstacle to a proof of Cardy's Formula for any other system is to circumvent or modify appropriately the color switching property.  The tack of this paper is along the latter course.  For our model we define a stochastic class of events known as \emph{path designates} and we meticulously enforce detailed criteria for which paths are to be considered.  It turns out that this requires the introduction of a host of auxiliary random variables which provide ``permissions'' for exceptions to the usual conventions of (self--avoiding) paths.   Furthermore, the random variables occasionally deny the existence of paths notwithstanding their appearance in the percolation configuration.  The end result is that a modified version of color switching symmetry is locally restored and an analogue of Smirnov's Cauchy--Riemann relations can be established.  Thereafter we can use a nearly identical contour--based argument to prove convergence of $u_{_N}$, $v_{_N}$ and $w_{_N}$ to the limiting $h$'s.   

\section{\large{Bond--Triangular Lattice Problems}}
\subsection{Preliminary Discussion}

We start with a brief recapitulation of the perspective on the usual bond-triangular lattice problems that was introduced in \cite{CL}.  Normally one considers the model where edges of the triangular lattice are independently declared to be occupied with probability $\lambda \in (0,1)$ and otherwise -- with probability $(1-\lambda)$ -- they are vacant.  Typically, the problems of interest are concerned with sets of sites connected by occupied bonds; paying heed only to the induced connectivity properties of the underlying sites, it  is clear that the bond description provides more information than is actually needed.  Indeed, focusing attention on a single triangle we see that out of the grande total of eight possible occupied/vacant edge configurations, there are only five distinguished outcomes: all sites connected, a pair of sites connected (which has three distinctive instances) and none of the sites connected.  

Thus, as far as percolation problems are concerned, we might as well just consider the problem where these five configurations are all that can be exhibited on a given triangle.  Furthermore, the structure of the full lattice allows the partition of the underlying space into disjoint triangles, e.g.~the up--pointing triangles, wherein each triangle independently exhibits one of the above mentioned five configurations.  

\begin{wrapfigure}{l}{0.51 \textwidth}
\vspace{-3mm}
\includegraphics[width=0.50 \textwidth]{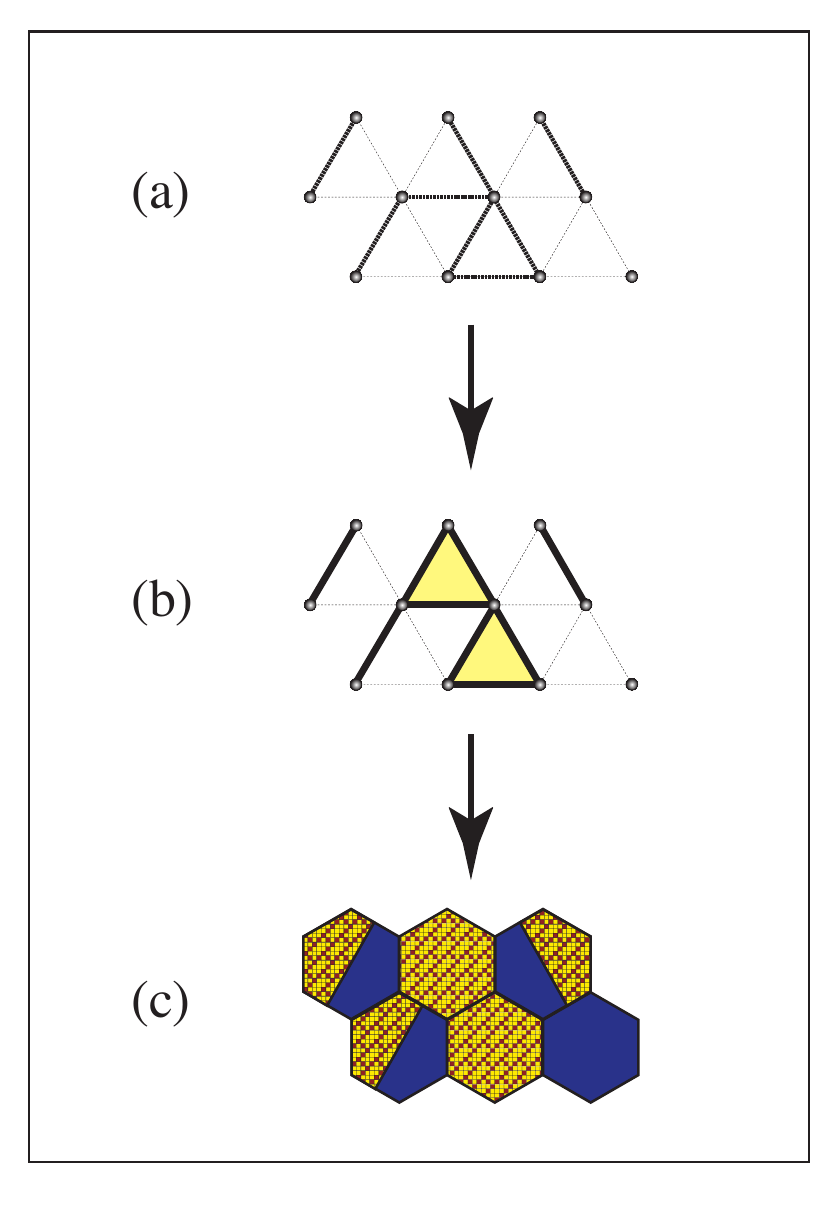}
\caption{\label{hex_f}\footnotesize{
Bond percolation as a hexagon tiling problem: (a) Typical bond configuration on the triangular lattice.  (b) Amalgamation into relevant connected objects.  (c) Associated tiling problem using hexagons and split hexagons.}}
\end{wrapfigure}

Needless to say, the configurations may still be represented by occupied and vacant bonds but, on up--pointing triangles, the original event of exactly two occupied bonds is identified with the full (three--bond) configuration.  From this perspective, it is natural -- and actually helpful -- to consider the general problem where the Bernoulli parameters are {\emph not} entangled by an underlying independent bond structure.  Thus we assign probabilities $a$ for all--bond event, $e$ for the empty event and $s$ for the three singles; $a + e + 3s = 1$.  It is noted that in the context of the $q$--state Potts model and the random cluster model of which this is the $q=1$ version, this enlargement of the problem amounts to the addition of three--body interactions in the Hamiltonian.  Under the star--triangle transformation, up--pointing triangles are replaced by superimposed down--pointing triangles and the parameters $a$ and $e$ get swapped, at least for $q=1$.  For more details see \cite{CL}.  But of immediate relevance to the subject of \emph{site} percolation on the triangular lattice (and all of its associated advantageous attributes) is the observation that for $s=0$, the above model on up--pointing triangles \emph{is} this site model with triangles playing the r\^ole of the sites.

As far as the present work is concerned, the crucial benefit of this ``packaged triangular'' description is the realization of these problems vis--\`a--vis hexagonal tilings.  Starting at the $s=0$ limit -- the site model -- we may replace each up--pointing (and/or superimposed, dual, down--pointing $*$--) triangle with a hexagon.  The hexagons tile the plane and, as is well--known from the site triangular model, exhibit the correct neighborhood connectivity relations, where, of course, connectivity is defined by the sharing of an edge.  

\begin{figure}[t]
\vspace{-3mm}
\begin{center}
\includegraphics[width=0.9 \textwidth]{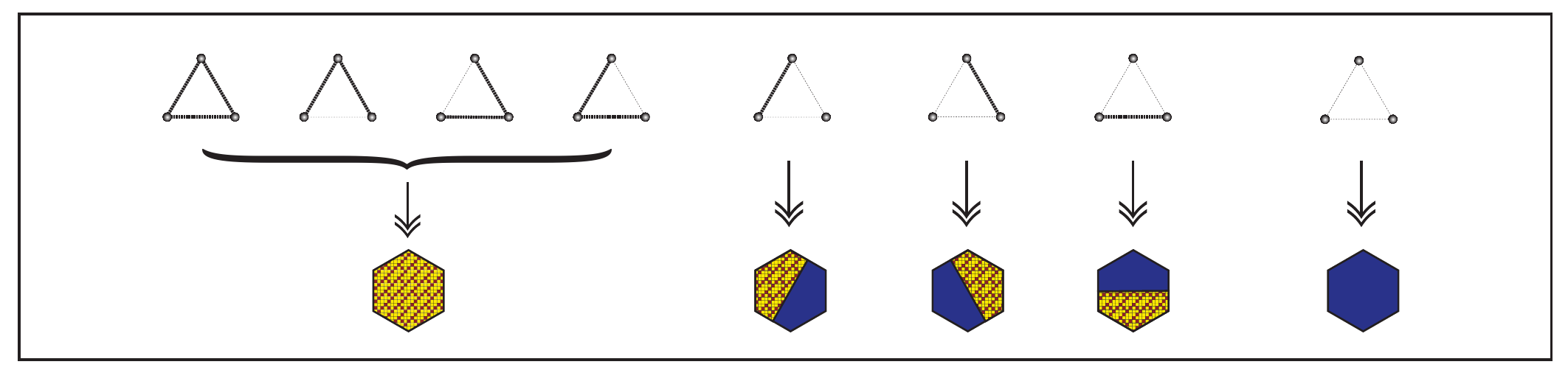}
\vspace{-5mm}
\end{center}
\caption{\label{hex_f2}\footnotesize{
Correspondence between eight configurations on (up--pointing) triangles and five hexagon configurations.  All four configurations which fully connect the triangle map to the single, fully yellow, hexagon with total weight $a$.  Empty configuration has probability $e$ and maps to the fully blue hexagon.  The three single bond configurations lead to split hexagons, each carrying probability $s$.  Note that not all the possible ways of splitting a hexagon appears:  Images obtained from the above three by reflections in the $x$--axis are not present.}}
\end{figure}

The bond model and its dual are now represented by a tile coloring problem: we color the hexagon blue if the corresponding up--pointing triangle is empty and yellow if it is all--bonds.   Yellow connectivity in the hexagon language corresponds to bond connectivity in the direct model while the connections between blue hexagons designate the connectivity properties of the dual model.  

As it turns out, a representation along these lines remains valid for $s > 0$.  We map single bond events associated with the original bond problem into hexagons that have been split along the diagonals connecting the midpoints of opposing edges and coloring them half--yellow and half--blue.  It is easy to check that this can be done in a consistent fashion so that the single bond events are faithfully represented, where two hexagons are now considered connected if they share either a full edge or half an edge (see Figure \ref{hex_f}).

 A few remarks on symmetry are in order.  First we note that only three of the six possible split hexagons occur.  This restriction breaks (microscopic) color symmetry for the models under consideration (see Figure \ref{hex_f2}).  The tiling model with all six split hexagons present (which enjoys full yellow--blue symmetry) can presumably be handled by a direct extension of \cite{Smir} but does not correspond to any realistic scenario in the language of the bond model.  Nevertheless, the set of three split hexagons do enjoy some symmetry of another sort: if we orient the hexagons so that two of the edges are parallel to the $y$--axis (as in all the figures) then the restricted set of three split hexagons does enjoy a reflection symmetry through the $y$--axis as well as the two axes at $\pm120^\circ$ to the $y$--axis.
As far as the $x$--axis and the other two axes are concerned, there is the more restrictive symmetry of reflection followed by color reversal.  


\subsection{\large{Setup, Definitions and the Model}}\label{def}  
We begin with a (more formal) recapitulation of the generalized triangular bond lattice problem in the hexagonal language, as it forms the basis of the model we will eventually study.  Consider a hexagonal tiling of the plane; to be definitive, the hexagons are oriented so that two of the edges are parallel to the $y$--axis.  With reference to the underlying bond model, the direct model will consist of up--pointing triangles and hence the superimposed down--pointing triangles constitute the ``dual'' lattice under the star--triangle transformation.  The color yellow will correspond to the direct model and blue to the dual model.  We call a hexagon which has only one color \emph{pure} and we call a hexagon which has two colors \emph{mixed}; the allowed mixed configurations are illustrated in Figure \ref{hex_f2}.

Using the hexagonal representation described in the last subsection, let $a$, $s$ and $e$ (with $a+e+3s=1$) denote the probabilities that a hexagon is pure yellow, mixed (one of three ways) or pure blue.  Occasionally, for the sake of clarity, we will use $y$ and $b$ instead of $a$ and $e$, which allows for effective tracking of various terms in up and coming formulae.  On general grounds \cite{CL}, the critical condition is simply $a=e$, which, as far as the pure hexagons are concerned, is the point of yellow--blue symmetry.  The usual independent bond model is just the curve in the $a$--$e$ plane $a = \lambda^3 + 3 \lambda^2 (1-\lambda)$, $e = (1-\lambda)^3$; where this curve hits the line $a = e$ is the star--triangle point.  We point out that this means for each value of $a = e$, we have a one parameter family, parametrized by $s$, of critical percolation models.  However, this is not the full story.  It turns out that we can appeal to FKG type inequalities (positive correlations) if and only if $ae \geq 2s^2$ \cite{CL} and, since this will prove necessary on occasion, we restrict ourselves to this range of parameters.

\begin{wrapfigure}{r}{0.40 \textwidth}
\includegraphics[width=0.40 \textwidth]{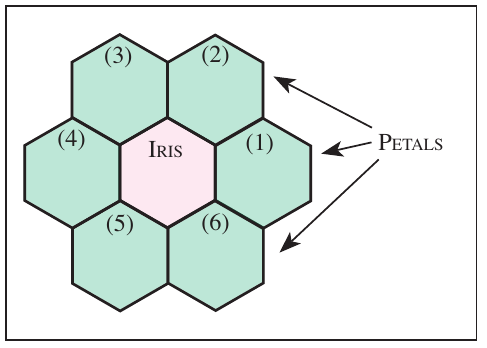}
\caption{\label{flower_fig}\footnotesize{
A flower.}}
\end{wrapfigure}

The full problem as described is, unfortunately, beyond our present capabilities.  In this paper, we will study a one parameter family of models which is on the one hand simpler than the full bond triangular lattice problem but on the other hand highlights some of the difficulties one encounters extracting continuum limits on lattices other than the triangular site lattice.  Our model is derived from the above by limiting the set of hexagons that are allowed to exhibit mixed configurations and introducing yet more local correlations.  Specifically, our efforts are focused on specific local arrangements of hexagons which we now describe.    

\begin{defn} We define a \emph{flower} to be a hexagon together with its six neighboring hexagons.  The central hexagon is called the \emph{iris} and the outer hexagons are called the \emph{petals} which are labeled 1 through 6 (and occasionally designated by other integers modulo 6), starting from the one directly to the right of the iris.  See Figure \ref{flower_fig}.
\end{defn}

For technical -- and complicated -- reasons, this work will be limited by restrictions on which hexagons are (and under what circumstances a hexagon is) allowed to exhibit the mixed states.  In particular, we envision a number of irises, whose flowers are disjoint, together with a background of \emph{filler sites}.  It is only the irises of the flowers which are allowed to exhibit the mixed hexagons.  In infinite volume we ultimately require the placement of the irises to have a periodic structure with $60^\circ$ symmetries, but we will not invoke this proviso till considerably later on.  For finite volumes, the specifics are as follows.

\begin{defn}\label{floral}
Consider a domain $\Lambda \subset \mathbb{C}$ which is tiled by hexagons and which we assume, for once and all, to be simply connected.  We identify $\Lambda$ with the set of hexagons tiling it. We say that $\Lambda_{\frak F}$ is a \emph{floral arrangement} of $\Lambda$ if certain designated hexagons of $\Lambda$, the irises, satisfy the following two criteria:\\
\indent $\bullet$ No iris is a boundary hexagon of $\Lambda$.\\
\indent $\bullet$ There are at least two non--iris hexagons between  each pair of irises.\\
Note that this means that the flowers associated with each iris are disjoint and are not ``broken across'' the boundary of $\Lambda$.
\end{defn}

\begin{wrapfigure}{r}{0.40 \textwidth}
\vspace{3 mm}
\includegraphics[width=0.40 \textwidth]{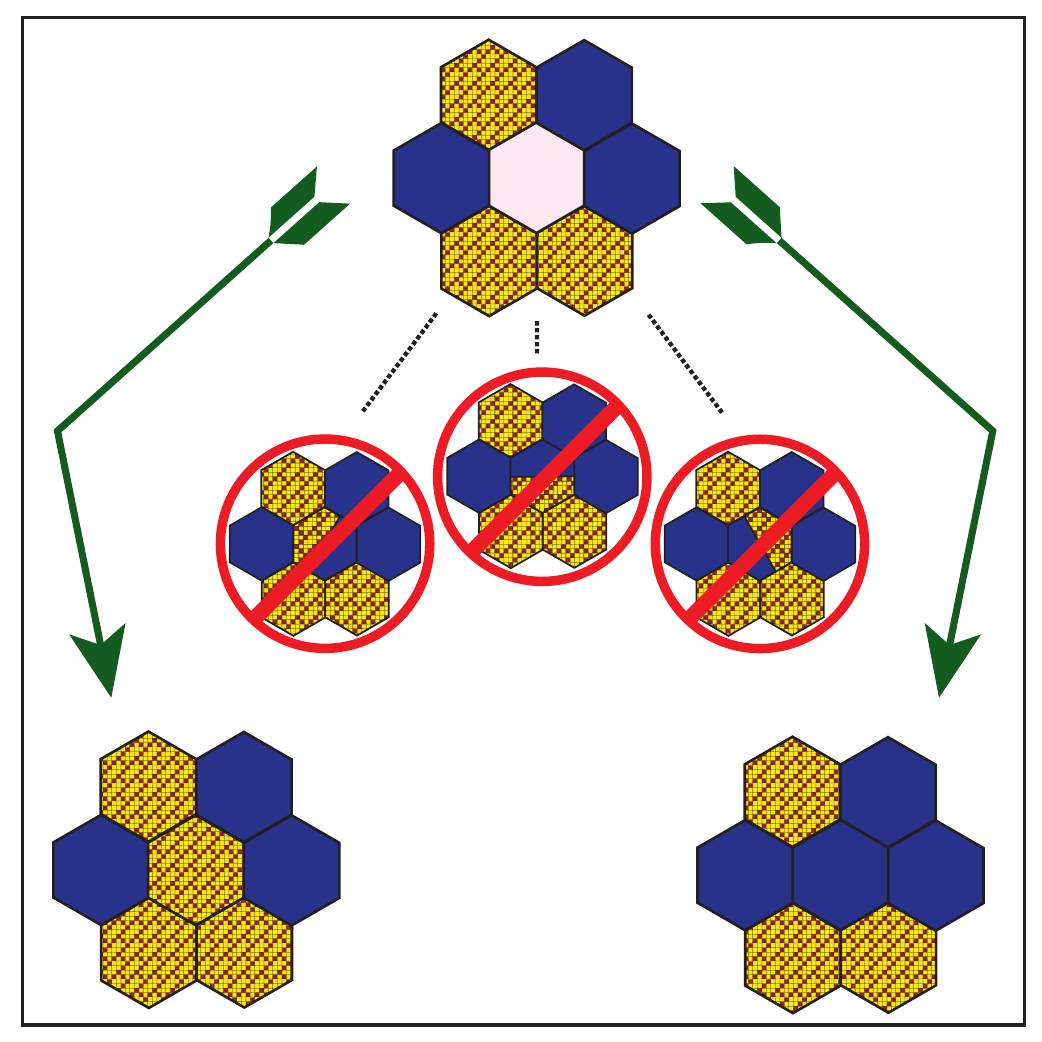}
\caption{\label{trig_fig}\footnotesize{
In a triggering configuration (three yellows, two of which are contiguous) a split hexagon is forbidden.  The iris is pure yellow or pure blue with conditional probabilities one--half.}}
\vspace{-10 mm}
\end{wrapfigure}

We now give a general description of our model:

\begin{defn}\label{model} Let $\Lambda$ be a domain with floral arrangement $\Lambda_{\frak{F}}$. 
\begin{itemize}
\item Any background filler sites, as well as the petal sites, must be $Y$ (pure yellow) or $B$ (pure blue), each with probability $\frac{1}{2}$.  In most configurations of the petals, we allow each iris to exhibit one of five states: $Y$, $B$, or the three mixed states $\alpha$ (horizontal split), $\beta$ ($120^\circ$ split) and $\gamma$ ($60^\circ$ split). Each mixed state occurs with probability $s$ and each pure state with probability $a = \frac{1}{2} (1-3s)$.
\item The exceptional configurations, which we call \emph{triggers}, are configurations where there are three yellow petals and three blue petals with exactly one pair of yellow (and hence one pair of blue) petals contiguous.  Under these circumstances, the iris is restricted to a pure form, i.e., blue or yellow with probability $\frac{1}{2}$.
\end{itemize}
All petal arrangements are independent, all flowers are configured independently, and these in turn are independent of the background filler sites (if any).  The resulting measure on these hexagon configurations will be denoted by $\mu$.
\end{defn}


For fixed $\Lambda_{\frak F}$, a \emph{configuration} $\omega \in \Omega_{\Lambda_{\frak F}}$ is an assignment of yellow or blue to all the petals in $\Lambda_{\frak F}$ and an assignment of one of the five types to each iris, in accordance with Definition \ref{model}.  Connectivity in $\omega$ is defined in the natural fashion; specifically, the notion of e.g.~blue connectivity may be defined as the usual $\mathbb R^2$ connectivity of (the closure of) the region that has been colored blue.  

\subsection{Scaling Limit and Statement of Main Theorem}

Percolation in our model is defined by considering a sequence of floral arrangements 
\[\Lambda_{\frak F_1}^{(1)}, \dots, \Lambda_{\frak F_k}^{(k)}, \dots \] with $\Lambda^{(j)} \subset \Lambda^{(j+1)}$; $\Lambda^{(j)} \nearrow \mathbb C$ and the $\Lambda_{\frak F_k}$'s \emph{consistent} in the sense that all the irises of $\Lambda_{\frak F_j}$ are in $\Lambda_{\frak F_{j+1}}$.  Then (pertinent to the extended model with differing parameters for pure blue and pure yellow hexagons) we say there is percolation of yellow's if some fixed point belongs to an infinite cluster of yellow with positive probability and similarly for blue's.  However, not surprisingly, it turns out that the model under consideration has no percolation (here is one instance in which we are forced to invoke our $60^\circ$ symmetry) and, as we will later demonstrate, the model exhibits all the well--known properties which are indicative of criticality in a 2D percolation problem (Theorem \ref{critical}).

To state our main result we need to introduce some minimal notation (more details to come in Section \ref{conv_proof}) and describe how the scaling limit is taken.  Let $\mathcal D \subset \mathbb C$ denote a domain with piecewise smooth boundary which is conformally equivalent to a triangle.  The boundaries and relevant prime ends will be denoted by $\mathscr A, \dots, e_{_{BC}}$.  We let $\tilde{\Lambda}_{\mathfrak F_N}$ denote an approximate discretization of $\mathcal D$ with lattice spacing $N^{-1}$ in accord with Definition \ref{floral}.  The version of $\tilde{\Lambda}_{\mathfrak F_N}$ rescaled to unit size will be denoted by $\Lambda_{\mathfrak F_N}$.  It is required that the $\Lambda_{\mathfrak F_N}$'s are consistent in the fashion described above.  The limiting floral arrangement will be denoted by $\Lambda_{\mathfrak F_\infty}$.

We write $z \in \Lambda_{\mathfrak F_N}$ if $z$ is a vertex of a hexagon in $\Lambda_{\mathfrak F_N}$.  For $z \in \Lambda_{\mathfrak F_N}$ we define the discrete function $\mathscrb U_{N}^B(z)$ to be the indicator function of the event that there is a blue path connecting the $\mathscr A$ and $\mathscr B$ boundaries which separates $z$ from $\mathscr C$.  We let $u_{_N}^B(z) = \mathbb E(\mathscrb U_N^B(z))$, with similar definitions for $v$ and $w$ and yellow paths.  We extend these functions in some suitable fashion off the lattice sites.  Then for $\mathscrb Z \in \mathcal D$ (unscaled), define $U_N^B(\mathscrb Z) = u_{_N}^B(Nz)$.

Our main result, convergence to the Cardy--Carleson functions, can now be stated: 

\begin{thm}\label{main theorem}
For the model as defined in Definition \ref{model}, with setup and notation as just described, under the conditions that
\[ a^2 \geq 2s^2\]
and
that $\Lambda_{\mathfrak F_\infty}$ is periodic and has $60^\circ$ symmetry,
we have 
\[\lim_{N \rightarrow \infty} U_N^B = h_{\mathscr C},\]
with similar results for $V_N^B$ and $W_N^B$.  The yellow versions of all of these functions converge to the same corresponding limiting functions.
\end{thm}

The key to all these considerations are the long--distance and local connectivity properties of the model.  This subject, along with the necessary deviations from the usual percolation scenarios is the content of the forthcoming section.

\section{Paths and Path Designates}
\subsection{Paths}
We start with a description of the paths we will be considering.  First we give a general definition for the usual notions of an allowed path and then describe exceptions in particular cases.  Under normal circumstances, a \emph{path} is a sequence of hexagons $(h_1, \dots, h_M)$ where $h_k$ and $h_j$ are neighbors (sharing an edge in common) if $|j-k| = 1$.  Additional rules may be implemented concerning \emph{hexagon self--avoidance}, i.e.~forbidding multiple usage of the same hexagon ($h_1, \dots, h_M$ are all distinct) and \emph{close encounters} ($h_k$ and $h_j$ neighbors with $|j-k| > 1$).  In most circumstances these supplementary conditions are immaterial; if there is a ``path'' from $h_1$ to $h_M$ with close encounters and multiple hexagon usage then there is a subsequence of these hexagons which forms the requisite path with neither close encounter nor multiple hexagon usage.  In this work, we will make use of all these phenotypes.  However, in various circumstances, it will be necessary that our paths represent {\it cuts}.  Thus we do not consider a sequence of hexagons $(h_1, \dots, h_M)$ to constitute a path unless successive interfaces between adjacent hexagons can be joined by a finite number of straight line segments which (in the continuum) culminate in a non--self--crossing path.  In particular, if the collection $\{h_1, \dots, h_M\}$ has the appearance of a path with a loop, one ordering is permitted, while the other -- which would force the straight line segments to cross -- is not considered legitimate.

Hence in any configuration of pure hexagons, there are blue and yellow paths.  With the injection of mixed hexagons into the picture, the necessary modifications are obvious; note the proviso that in a colored path with mixed elements, the relevant portions of successive hexagons are required to share at least \emph{half} an edge in common.  More precisely, here is a definition.

\begin{defn}
Let $(h_1, \dots, h_M)$ denote a path and $\omega$ a configuration in some $\Lambda_{\frak F}$.  We will say that the path is a \emph{blue transmit} in $\omega$ if each of $h_1, \dots, h_M$ is either pure blue or, if $h_j$ is mixed, the blue part of $h_j$ shares at least half an edge with both $h_{j-1}$ and $h_{j+1}$ and thereby connects $h_{j-1}$ to $h_{j+1}$.  Similarly for a \emph{yellow transmit}.
\end{defn}    
	
Typically -- as was evidently the case in \cite{Smir} -- on any path, multiple usage is forbidden and close encounters are indulged.  We remark that these normally inconsequential provisos are only slightly important in the definition of the events $\mathscrb U_{_N}(z)$, $\mathscrb V_{_N}(z)$ and $\mathscrb W_{_N}(z)$ (cf. Definition \ref{functions}), but they become essential when it comes to the derivatives of their probabilities.  In particular, as to the definitions of the paths satisfying these events we will occasionally forbid touches and (as sort of a compensation) we will occasionally allow multiple usage.  These exceptions will be stochastically implemented according to the details of the local configuration.  

\begin{remark}\label{self--avoiding?}
We remark that there are certain self--avoiding paths which, by the standards of the pure model, would not be called self--avoiding.  Indeed, consider a horizontal blue transmit across a flower with the iris in the $\alpha$--state (horizontal split, blue on top).  If the next hexagon in the path sequence is petal 6, so that the sequence is now [4; iris; 1; 6], the path has the appearance of a redundant visit to petal 1.  However, due to the mixed nature of the iris, it is seen that in fact all the hexagons specified are necessary for the connection between petal 4 and petal 1.  The preceding example illustrates that it is just the blue parts of the path that have to be self--avoiding, which is a property directly inherited from the ``correct'' notions of self--avoiding in the underlying bond model.  These phenomena lead to some interesting scenarios whereby the geometric structure of a self--avoiding path sometimes does and sometimes does not reveal the underlying state of the iris.
\end{remark}

\subsection{Path Designates}
A key technical device in this work is to replace the usual (i.e.~full) description of paths with partial information to arrive at a set of objects called \emph{path designates}.  By the usual abuse of notation, we will use the phrase path designate to describe both events and geometric objects.  With regards to the latter a path designate is, for all intents and purposes, a collection of paths.  So, for pedagogical purposes, let us start with a microscopic path and describe which path designate it belongs to.  Consider the portion of the path that intersects a particular flower.  In the simplest case, the path only visits the flower once and thus there is an \emph{entrance petal} and an \emph{exit petal}.  In contrast to the microscopic description where it must be specified how the path got between these ``ports'', we leave these details unsaid.  Similarly, with multiple visits to a single flower, the first entrance and exit petals, the second entrance and exit petals, 
 etc.~must all be specified.  This must be done for \emph{all} flowers and on the region complementary to the flowers (if any) the path must be entirely specified.  Note that, with only slight loss of generality, path designates do not begin or end on irises.  A formal definition is as follows:  

\begin{defn} (Path Designate)
Let $\Lambda_{\frak F}$ denote a floral arrangement.  A \emph{path designate} in $\Lambda$ from $h_0$ to $h_{K+1}$ is a sequence
\begin{equation*}
[H_{0, 1}, (\mathscr{F}_1, h_1^e, h_1^x), H_{1, 2}, (\mathscr{F}_2, h_2^e, h_2^x), H_{2, 3}, \dots, (\mathscr{F}_K, h_K^e, h_K^x), H_{K, K+1}]
\end{equation*}
where $\mathscr F_1, \dots, \mathscr F_K$ are flowers in $\Lambda_{\frak F}$, $h_j^e$ and $h_j^x$ are (entrance and exit) petals in the $j^{th}$ flower and, for $1 \leq j \leq K-1$, $H_{j, j+1}$ is a path in the complement of flowers which connects $h_j^x$ to $h_{j+1}^e$.  Further, $H_{01}$ is a path in the complement of flowers from $h_0$ to $h_1^e$ and similarly $H_{K, K+1}$ is a path from $h_K^x$ to $h_{K+1}$ in the complement of flowers.  We note that in the above definition, not all flowers have to be distinct: $h_j^e$ could equal $h_j^x$ -- i.e.~the flower is visited at a single petal and, depending on the floral arrangement, the $H_{j, j+1}$'s could be vacuous.  However, we shall assume, with negligible loss of generality, that all of the explicitly mentioned hexagons  (i.e., the collection of hexagons which constitute the paths $H_{j, j+1}$ along with the entrance and exit hexagons) in a path designate are used only once.
\end{defn}

Of course, for percolation problems the only matter of importance is the realization of underlying paths.  Thus the following is obviously relevant:  

\begin{defn} (Realization of a Path Designate)
Let $\mathscr P$ denote a path designate.  We let $\mathscr P_B$ denote the event that for all $j$, all hexagons in the path $H_{j, j+1}$ as well as $h^e_j$ and $h^x_j$ are blue and there is a blue connection in $\mathfrak F_j$ between $h^e_j$ and $h^x_j$.  A similar definition holds for the event $\mathscr P_Y$.
\end{defn}

\begin{remark}
Clearly the event $\mathscr P_B$ means that the designate $\mathscr P$ is ``achieved'' (or ``transmitted'') by an underlying blue path.  However, there is no guarantee that the underlying blue path has reasonable self--avoidance properties.  Indeed, it may be the case that the path is inundated with close encounters;  in particular, entrance and exit hexagons may be used in a seemingly redundant way.  These matters will be of no concern and in our derivations we will be dealing exclusively with path designates and the events that various transmissions along these designates are achieved.
\end{remark}

We begin with a preliminary demonstration of how the path designates might allow us to implement microscopic color switching.  In particular, and of seminal importance for the present model, is the following:

\begin{lemma}\label{unconditioned}
Let $\Lambda_{\mathfrak F}$ denote a floral arrangement and let $\mathbf{r}$, $\mathbf{r}^\prime$ denote points (hexagons) in $\Lambda_{\mathfrak F}$ which are not irises.  Let $K_{\mathbf r \mathbf r^\prime}^{B}$ denote the event of a blue transmission between $\mathbf{r}$ and $\mathbf{r}^\prime$, and similarly for $K_{\mathbf r \mathbf r^\prime}^{Y}$.  Consider the model as described in Definition \ref{model} and let $\kappa_{\mathbf r \mathbf r^\prime}^B = \mathbb{P}(K_{\mathbf r \mathbf r^\prime}^B)$ with a similar definition for $\kappa_{\mathbf r \mathbf r^\prime}^Y$.  Then
\begin{equation*} \kappa_{\mathbf r \mathbf r^\prime}^B = \kappa_{\mathbf r \mathbf r^\prime}^Y. \end{equation*}
\end{lemma}

Before the proof of Lemma \ref{unconditioned} we will need a preliminary lemma, and, of course, some further definitions.

\begin{defn}
Let $\mathscr F$ denote a flower and $\mathscr D$ a collection of petals.  Let $T^B_{\mathscr D}$ denote the event that all the petals in $\mathscr D$ are blue and that they are blue connected within the flower.  Let $T^Y_{\mathscr D}$ denote a similar event with blue replaced by yellow.   
\end{defn}

\begin{lemma}\label{single collection}
For all $\mathscr D$, 
\begin{equation*} \mathbb{P}(T^B_{\mathscr D}) = \mathbb{P}(T^Y_{\mathscr D}).
\end{equation*}
\end{lemma}

\noindent
{\bf Proof: }Let $\eta$ denote a configuration on the petals and $\overline{\eta}$ the color reverse of $\eta$.  Clearly, it is enough to show that (for all $\eta$)
\begin{equation*}
\mathbb P(T_{\mathscr D}^B \mid \eta) = \mathbb P(T_{\mathscr D}^Y \mid \overline{\eta}).
\end{equation*}
It may be assumed without further discussion that all petals in $\mathscr D$ are already blue in $\eta$ (otherwise both sides of the previous equation are zero).  If $\mathscr D$ is already blue connected in $\eta$ then there is nothing to prove.  If $\eta$ is a trigger, then there is also nothing to prove because of full color symmetry.  In general $\mathscr D$ cannot have more than three components.  In the case of three, if none of these have been connected in $\eta$ then the only possibility is the alternating configuration which, as can easily be checked, requires a pure iris to achieve full connectivity.  We are thus down to two separate components in $\eta$ which need to be connected through the iris.  

To be specific, let us study the blue version of this problem.  For all intents and purposes, the only cases that need be considered are the ones where $\eta$ has two non--adjacent blue petals (which need to be connected through the iris) and all other petals yellow.  Now, it turns out that either the blue petals are blue connected through the iris or the complementary ``yellow'' sets are yellow connected through the iris -- a micro--environment duality.  To dispense with the present case, we invoke (and not for the last time) the fact that for two non--adjacent petals of the same color, there is one and only one mixed hexagon which permits the successful transmission of their color.  Thus, for all the cases where $\eta$ has exactly two usable blue petals we have 
\begin{equation*}
\mathbb{P}(T_{\mathscr D}^B \mid \eta) = b + s
\end{equation*}
with a similar result for $\mathbb{P}(T_{\mathscr D}^Y \mid \overline{\eta})$.  But now, by the above--mentioned duality, any other (non--trigger) two--component case which involves more than just two usable petals of the same color has probability given by $a+2s$.  
\qed

\begin{remark}\label{multi}
We will, formally, have to consider cases involving several sets; e.g., $\mathscr D_1$, $\mathscr D_2$, $\dots$, $\mathscr D_k$ and $T^Y_{\mathscr D_1\dots \mathscr D_k}$, the event that all the relevant $\mathscr D's$ are yellow connected sets, but not necessarily all connected to each other.  Due to the limitations of the flower size, it is seen that any case with $k \geq 3$ is trivial or reduces to $k < 3$.  The only non--trivial case with $k = 2$ is exemplified by the problem where $\mathscr D_1$ consists of two petals separated by another petal and $\mathscr D_2$ a single petal separated from both of these by yet another petal -- the alternating configuration.  Here either $\eta$ reduces this back to a single--$\mathscr D$ problem or, if all the other petals are blue, the desired result (transmission color symmetry) follows from the previous observation that each binary transmission through the iris is permitted by exactly one mixed hexagon for both yellow and blue.  We therefore 
 consider the multi--set version of Lemma \ref{single collection} to be proved.
\end{remark}

\noindent
{\bf Proof of Lemma \ref{unconditioned}: }Let $\mathbf{r}$ and $\mathbf{r}^\prime$ denote two non--iris points in $\Lambda_{\mathfrak F}$.  We first observe that the event of a blue transmission between $\mathbf{r}$ and $\mathbf{r}^\prime$ is also the event that there exists a $\mathscr P$ beginning at $\mathbf{r}$ and ending at $\mathbf{r}^\prime$ such that $\mathscr P_B$ occurs.  In particular, letting $\Pi_{\mathbf r \mathbf r^\prime}$ denote the collection of all path designates beginning at $\mathbf{r}$ and ending at $\mathbf{r}^\prime$, we have
\begin{equation}\label{PtoD}
\kappa_{\mathbf r \mathbf r^\prime}^B = \mathbb{P}(\bigcup_{\mathscr P\in\Pi_{\mathbf r \mathbf r^\prime}} \mathscr P_B) 
\end{equation}
and similarly for $\kappa_{\mathbf r \mathbf r^\prime}^Y$.  Noting that $|\Pi_{\mathbf r \mathbf r^\prime}| < \infty$, we will handle the likes of (\ref{PtoD}) via an inclusion--exclusion argument.  Let us first demonstrate that for any $\mathscr P$, 
\begin{equation*}
\mathbb{P}(\mathscr P_Y)=\mathbb{P}(\mathscr P_B).
\end{equation*}
Indeed, we write 
\begin{equation*}
\mathscr P = [H_{\mathbf r 1}, (\mathscr{F}_1, h_1^e, h_1^x), H_{12}, (\mathscr{F}_2, h_2^e, h_2^x), H_{23}, \dots, (\mathscr{F}_K, h_K^e, h_K^x), H_{K\mathbf r^\prime}],
\end{equation*}
where $\mathbf r$ is used to denote the hexagon at $\mathbf r$, etc.  Assuming for simplicity that each flower is used only once, the formula for $\mathbb{P}(\mathscr P_B)$ is given by the product along successive terms:
\begin{equation*}
\mathbb P(\mathscr P_B) = \left(\frac{1}{2}\right)^{|H_{\mathbf r, 1}|}\mathbb P(T^B_{\{h^e_1, h^x_1\}}) \left(\frac{1}{2}\right)^{|H_{1, 2}|} \dots \hspace{3mm} \mathbb P(T^B_{\{h^e_k, h^x_k\}}) \left(\frac{1}{2}\right)^{|H_{K, \mathbf r^\prime}|}.
\end{equation*}
By Lemma \ref{single collection}, all terms are the same when $B$ is replaced by $Y$.  In more generality -- for the case of a single path -- various pairs or triples of transmission terms which actually involve the same flower must be treated in one piece.  E.g., if $\mathfrak F_\ell = \mathfrak F_j$ and, say, $h^e_\ell = h^x_\ell$ while $h^e_j \neq h^x_j$ which is in turn distinct from $h^e_\ell$, then we would replace $\mathbb P(T^B_{\{h^e_\ell\}}) \mathbb P(T^B_{\{h^e_j, h^x_j\}})$ by $\mathbb P(T^B_{\{h^e_\ell\}, \{h^e_j, h^x_j\}})$.  In any case, by Lemma \ref{single collection} and Remark \ref{multi}, each term in the expression for blue transmission is equal to the corresponding term in the expression for yellow transmission. 

The general term in an inclusion--exclusion expansion will be of the form:
\begin{equation*}
\pm \mathbb P((\mathscr P_1)_B \cap (\mathscr P_2)_B \cap \dots \cap (\mathscr P_\ell)_B). 
\end{equation*} 
These terms will be handled in a manner similar to the single path case.  Indeed, first we will need an overall term representing the amalgamation of all the outside hexagons (if any); this will be $\frac{1}{2}$ to some power, which will be the same for yellow as for blue.  Then, for each flower which appears in any of the relevant designates, we will need to multiply in a blue transmission probability to ensure that all the relevant entrance hexagons are connected to their corresponding exit hexagons, i.e.~a term of the form $\mathbb P(T^B_{\mathscr D_1, \mathscr D_2, \dots, \mathscr D_m})$.  However, by Lemma \ref{single collection} and Remark \ref{multi}, these blue transmission probabilities are, once again, the same as they are for yellow.  Thus, down to the level of each term in inclusion--exclusion, we have equality and the lemma is proved.  
\qed

The preceding is entirely general provided the floral arrangement adheres to the criteria spelled out in Definitions \ref{floral} and \ref{model}.  We augment this with some additional stipulations in order to obtain: 

\begin{thm}\label{critical}
Consider the model as described in Definition \ref{model} with the periodicity and $60^\circ$ symmetry assumptions discussed in the paragraph prior to Definition \ref{floral} and with the additional proviso that $a^2 \geq 2s^2$.  Then the model exhibits all the typical properties of a $2D$ percolation model at criticality: 
\begin{itemize}
\item There is no percolation of either the blue or yellow connected clusters.
\item Crossings of squares and rectangles have probabilities uniformly bounded above and below independent of scale (but dependent on their aspect ratio).
\item In any annulus of the form $S_L \setminus S_{\lambda L}$, where $S_L$ is a square of scale $L$ centered at the origin and $\lambda \in (0,1)$ with probability  bounded uniformly (in $L$) above and below, there is a yellow ring and/or a blue ring separating the outer boundary of $S_L$ from the origin.  
\item The probability of a connection between a fixed site and any other site a distance $n$ away is bounded above by an inverse power of $n$.
\item The probability of a connection between two distant sites is bounded above and below by a power of their separation.  
\end{itemize}
\end{thm}
\noindent
{\bf Proof: }In essence all of the above follows from Russo--Seymore--Welsh (\cite{Russo}, \cite{SW}) type arguments, which extend a lower bound on the probability of short way crossings of rectangles to lower bounds on the probabilities of crossing \emph{longer} rectangles; of crucial importance will be the fact that the ultimate bounds are uniform in $L$.  For these types of arguments an essential ingredient is, ostensibly, the Harris--FKG property.  It turns out that full monotonicity properties for the measure do not hold, however, as will be proved later, Lemma \ref{FKG} in the Appendix, a restricted form of the Harris--FKG property holds for all paths and path type events.  This lemma is proved under the proviso that $a^2 \geq 2s^2$.  Thus as far as RSW lemmas are concerned, we are free to use these sorts of correlation inequalities.

In point of fact, we will not use the argument of either the above references, but will rely on the methods of Lemma 6.1 in \cite{KBook}.  A necessary input for Lemma 6.1 in \cite{KBook} is bounds on the crossing probabilities of rectangles with aspect ratios not terribly different from unity.  We start with the establishment of a uniform bound on the probability of ``easy'' way crossings of rectangles with an aspect ratio of approximately $2$\hspace{1mm}:\hspace{1mm}$\sqrt{3}$.  (We note here that due to the microscopic structure of the hexagonal lattice and the occasional necessity to cut out irises at the boundary, there will be rough edges to the rectangles and to other shapes which are to follow.  These and future similar issues are not terribly important and will not be mentioned explicitly.)  The following, we assume, is standard for models with $60^\circ$ symmetry: 

Consider a large hexagon, of scale $L$, which is oriented in the same way as in Section \ref{def}; i.e.~with a set of edges parallel to the $y$--axis.   Without loss of generality, we assume that $L$ is commensurate with the period of the tiling and that the vertical line which splits this hexagon in half is a line of symmetry for the model.  Let us discuss the event of a yellow connection between, say, the left edge and one of its second--neighbor edges.  Our first claim is that if this event has a probability of order unity independent of $L$, then any connection between any pair of edges has a similar sort of bound.  Indeed, by $120^\circ$ symmetry this is manifestly true for the triad of next--neighbor faces anchored on the left side, and the opposite triad follows from reflection symmetry through the $y$--axis.  It is not hard to see that when all second--neighbor edges have probabilities of order unity of being connected, then (here we use the Harris--FKG property) any pair of edges are connected with a probability of order unity.  However, once these probabilities are established for yellow, then by Lemma \ref{unconditioned}, the same holds for blue -- and vice versa. Thus let us proceed with the event in question.  

If this event fails, then at least one of two dual blue events of a similar type must occur and/or a blue connection between the appropriate pair of opposing edges.  In the former case, we are done by the above--mentioned color symmetry.  In the latter case  (blue success with opposing edges), by employing an $120^\circ$ symmetry and taking the intersection of two such opposing edge events, we get, by the Harris--FKG property, the desired sort of connection (albeit in blue).  Having established the preliminary claims, it turns out that all we have use for is a horizontal crossing between the opposing edges.  Inscribing the hexagon in a rectangle with the above stated aspect ratio, we are finished with the horizontal problem.

For the vertical problem we first reorient the big hexagon so that two edges are parallel to the $x$--axis.  We may now proceed in almost the identical fashion, except that whereas in the previous argument, we employed the simple symmetry of $y$--reflections, here we employ the reflection through the $x$--axis combined with color reversal, which, as mentioned earlier, is another inherent symmetry of the model.  However, after this spurious color reversal, we may restore the original color by appeal to Lemma \ref{unconditioned}.  

We have gathered the following ingredients as inputs for Lemma 6.1 in \cite{KBook}: lower bounds on vertical and horizontal crossing probabilities of suitable rectangles (the requisite aspect ratios must, as it turns out, have a product that is not in excess of $3/2$) Harris--FKG properties for paths, and symmetry with respect to reflections through lines parallel to the $y$--axis.  One may follow the steps in Lemma 6.1 of \cite{KBook}, modifying and abridging when appropriate.  

Once we have vertical and horizontal crossings of long rectangles, the establishment of power law bounds, rings in annuli, etc.~follow -- with the help of Harris--FKG properties -- standard 2D percolation arguments.  We remark that some of these properties (e.g.~the power law lower bounds) but unfortunately not the crucial ones, can be established without the benefit of the RSW lemmas.
\qed
 
\subsection{\large{Color--Switching Lemmas}}

In the previous subsection, where paths were free to wander throughout the relevant domain, complete parity between yellow and blue was established.  However, as can be gleaned from the introduction, it will be necessary to establish this sort of equivalence in the presence of pre--existing blue or yellow paths; e.g.~the probability of a yellow/blue path connecting a pair of sets $A$ and $B$ in the presence of -- and disjoint from -- other paths connecting other sets.  While there is no doubt of such parity in the long view (i.e.~in a statistical sense on a large length scale), on the microscopic scale, yellow--blue equivalence will break down, as the following example demonstrates.

\begin{wrapfigure}{l}{0.53 \textwidth}
\vspace{-3mm}
\includegraphics[width=0.52 \textwidth]{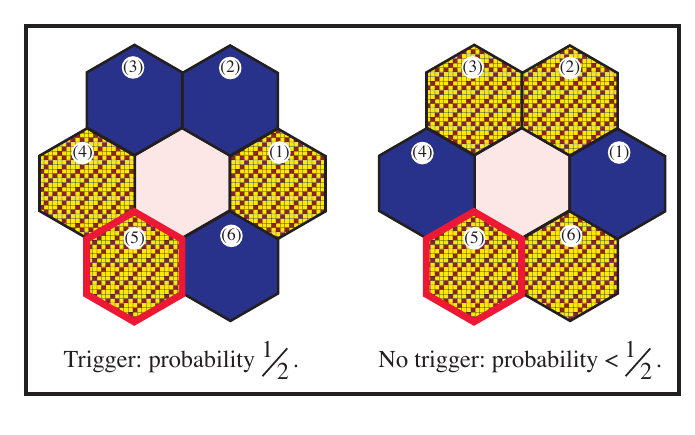}
\caption{\label{figure_trigger}\footnotesize{
A circumstance leading to asymmetry in conditional color switching.}}
\end{wrapfigure}  

\begin{ex}
As an example consider the probability that petals 2, 3 and 6 are connected in the complement of petal 5 -- which is conditioned to be yellow.  If the connection is achieved by going through the petals (without using 5) the yellow and blue transmission probabilities are the same.  However, on the transmissions through the iris, the probability of petals 2, 3 and 6 being blue and connected in the complement of petals 1, 4 and 5 (all of which are yellow) is $\frac{1}{2}$ since this is a triggering situation.  On the other hand, the situation with all petal colors (save the one that is conditioned, i.e.~petal 5) reversed gives that the probability of a yellow connection between 2, 3 and 6 is only $y+s < \frac{1}{2}$.    
\end{ex}

Our cure for these microscopic difficulties will be, in essence, to define away our problems.  Indeed, in the up and coming we will establish some results concerning transmissions through flowers with conditioned petals.  These transmissions are supposed to represent the construction of path segments in the presence of segments of other paths where all paths under consideration are meant to be disjoint.  We may therefore restore yellow--blue parity \emph{at the microscopic level} by relaxing the strict conventions which apply to disjoint paths.  In particular, while ``disjoint'' paths usually are interpreted as allowing the paths to touch while not sharing hexagons, here we will implement a special set of rules which permits some exceptions.  These will typically be denoted by a $*$, and the definition is as follows: 

\begin{defn}\label{diamond}
Let $\diamondsuit$ denote a configuration on a proper subset of the petals of a flower.  For $\mathscr D$ a set of petals (or a collection of sets of petals, c.f.~Remark \ref{multi}) on the complement of $\diamondsuit$, we consider the events $T_{\mathscr D, \diamondsuit}^B$ and $T_{\mathscr D, \diamondsuit}^Y$ defined by
\begin{equation*}
T_{\mathscr D, \diamondsuit}^B = \{\omega \mid \mathscr D \mbox{ is blue and all connected up in the complement of $\diamondsuit$}\},
\end{equation*}
and similarly for $T_{\mathscr D, \diamondsuit}^Y$.  The $*$\emph{--transmissions}, denoted by $T_{\mathscr D, \diamondsuit}^{B*}$ and $T_{\mathscr D, \diamondsuit}^{Y*}$ are events defined on a larger space.  Letting $\eta_\diamondsuit$ denote the full petal configuration, we have for each flower $\mathfrak F_k$ a collection $\mathscr X^k$ of 3--valued random variables $X^k_{\mathscr D, \diamondsuit} \in \{\textsc{o}, \textsc{y}, \textsc{b}\}$.  Focusing on a single flower, with $\mathscr D$ and $\diamondsuit$ fixed, and denoting the random variable by $X$ (notwithstanding that there are, literally, thousands of these objects), we have,
\begin{equation*}
\mbox{if $X=\textsc{o}$, then:\hspace{2mm} }
T_{\mathscr D, \diamondsuit}^{B*} = T_{\mathscr D, \diamondsuit}^B \mbox{  and } T_{\mathscr D, \diamondsuit}^{Y*} = T_{\mathscr D, \diamondsuit}^Y.
\end{equation*}
However, if $X= \textsc{b}$, then 
\begin{equation*}\begin{split}
T_{\mathscr D, \diamondsuit}^{B*} \cap \{X = \textsc{b}\}
= \{\omega \mid \mathscr D &\mbox{ is blue and all connected up}\\
&\mbox{ possibly using the blue petals of $\diamondsuit$\}} \cap \{X = \textsc{b}\}
\end{split}\end{equation*}
and
\begin{equation*}\begin{split}
T_{\mathscr D, \diamondsuit}^{Y*} \cap \{X =\textsc{b}\} = \{\omega \mid \mathscr D &\mbox{ is yellow and all connected up}\\
&\mbox{ without touching any yellow petals of $\diamondsuit$\}}\cap \{X=\textsc{b}\}. 
\end{split}\end{equation*}
Similar definitions hold for when $X = \textsc{y}$ with the roles of the transmission colors reversed.  We remind the reader that in case $\mathscr D$ refers to multiple sets, the connections need not be disjoint.  It is observed that for certain $\diamondsuit$ and $\mathscr D$, some of the above may be vacuous; this is an extreme case of a seminal point which will be exploited later.  We will call an assignment of these conditional probabilities (for the values of $X$) a set of $*$\emph{--rules} and the corresponding transmissions $*$--transmissions.
\end{defn}  

Our microscopic rebalancing will be broken down into two lemmas, ordered by conceptual difficulty.  The first deals exclusively with the cases where the iris is not involved in the conditioning and the second where it is.  The conceptual difference is that in the latter cases, the nature of the hexagon at the iris itself may change.  Fortunately, in these latter set of circumstances there are only a limited number of possibilities to consider. 

\begin{lemma}\label{noiris}
Let $\mathfrak F$ denote a flower and $\diamondsuit$ a partial configuration on the petals -- with all petals in $\diamondsuit$ being yellow.  Then for $X_{\mathscr D, \diamondsuit} \in \{\textsc{o}, \textsc{y}, \textsc{b}\}$, consider the $*$--transmissions $T_{\mathscr D, \diamondsuit}^{Y*}$ as defined in Definition \ref{diamond}.  Then there are joint laws for the $X_{\mathscr D, \diamondsuit}$'s such that 
\begin{equation*}
\mu(T_{\mathscr D, \diamondsuit}^{B}) = \mu^*(T_{\mathscr D, \diamondsuit}^{Y*}),
\end{equation*}
where $\mu^*$ denotes the joint probability measure on the flower configurations and $\mathscr X^k$ with marginal $\mu$.  Similar results hold with the role of yellow and blue reversed and, in case $\diamondsuit$ has petals of both colors, $*$--probabilities for the $*$--transmissions of the two colors are equal:
\begin{equation*}
\mu^*(T_{\mathscr D, \diamondsuit}^{B^*}) = \mu^*(T_{\mathscr D, \diamondsuit}^{Y*}),
\end{equation*}
\end{lemma}
\noindent
{\bf Proof: }We will in fact prove the stronger statement
\begin{equation}\label{bayes}
\mu^*(T_{\mathscr D, \diamondsuit}^{Y*} \mid \eta_\diamondsuit) = \mu(T_{\mathscr D, \diamondsuit}^B \mid \overline{\eta}_\diamondsuit),
\end{equation}
where $\overline{\eta}_\diamondsuit$ denotes the color reverse on the complement of $\diamondsuit$.  The above implies the desired result because the \emph{petal} configurations are provided by independent Bernoulli statistics.  We need not discuss trivial cases when the configuration of $\eta$ does not provide the necessary yellow petals of $\mathscr D$.  Furthermore, with the exception of a single configuration, i.e.~the alternating configuration, it turns out that without loss of generality, we may regard the yellow petals of $\eta$ that are contiguous to $\mathscr D$ as part of $\mathscr D$.

We therefore do a case by case analysis, starting with the situation where $\diamondsuit$ is but a single petal (which, without loss of generality, we have assumed to be yellow).  If on the complement of $\diamondsuit$ there are five yellow petals in $\eta_\diamondsuit$ then there is nothing to prove, and with four yellow petals, essentially nothing to prove.  Indeed, assuming those four petals are not contiguous, there is either the three and one split or the two and two split.  The desired result for the two and two split follows from symmetry (the blue petal \emph{must} be diametrically opposed to the conditioned petal which implies that the line joining them is an axis of reflection/color reverse symmetry).  The three and one splits follow similarly from this inherent reflection/color reverse symmetry.  E.g.~if the conditioned hexagon is petal 3 and the blue petals are at $\pm 1$, then transmission equality follows from the  symmetry of reflection through the $x$-axis followed by reversal of \emph{all} colors.    

\begin{wrapfigure}{l}{0.51 \textwidth}
\vspace{-3mm}
\includegraphics[width=0.50 \textwidth]{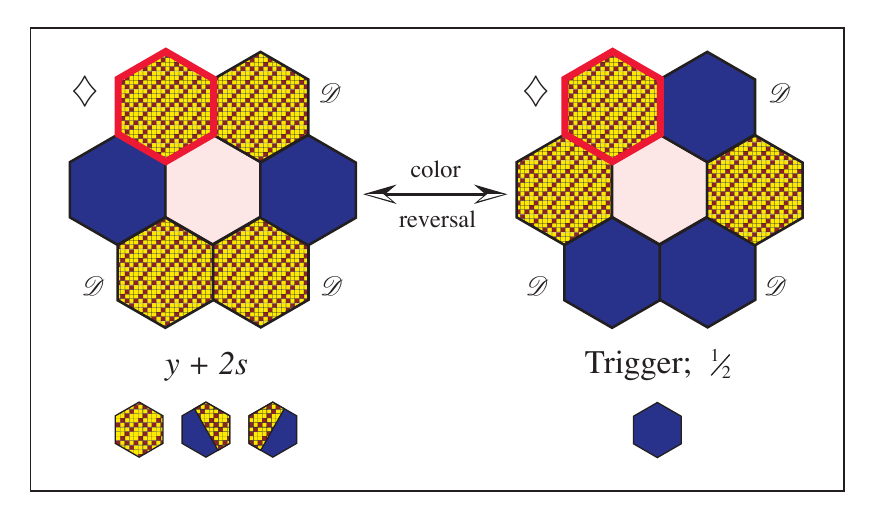}
\caption{\label{illustration}\footnotesize{A case with $|\diamond| = 1$ and $|\mathcal D| = 3$.}}
\end{wrapfigure}

The three petal cases -- those which are non--trivial -- are initially ominous looking, but can be easily handled with the added flexibility of implementing special rules.  First we discuss the more serious cases where two of the three petals are contiguous.  Whenever we have both the frozen petal and the pertinent trio in $\eta_\diamondsuit$ all yellow, triggers can only occur in the color reverse $\overline{\eta}_\diamondsuit$. Under these conditions, the relevant (conditional) blue transmission probabilities will all be $\frac{1}{2}$.  As for the yellow transmissions -- where there is no trigger -- the result will be either $y+s$ or $y+2s$, neither of which is $\frac{1}{2}$.  However, in the $y+s < \frac{1}{2}$ cases, where yellow would have the lower transmission probability, we may stochastically implement permission to share the conditioned petal.  As can be readily checked, since there are four (out of six) active petals in play, the extra petal is always in position to enhance the transmission probability.  Indeed, in certain cases, the implementation of the sharing automatically creates the desired connection and in the other cases it boosts the transmission probability  up to $y+2s > \frac{1}{2}$.  Thus, allowing sharing with the appropriate probability (e.g.~probability $\frac{1}{2}$ in the latter mentioned cases), we restore balance.  To deal with the cases where yellow has the {\it a priori} higher transmission probability,  first observe that since we have three yellow petals which are not contiguous, one of them must be adjacent to the conditioned petal.  We may therefore implement the rule forbidding close encounters with the appropriate probability, which happens to be $s/(2y + 4s)$.  This is illustrated in Figure \ref{illustration}.

Finally, to finish the cases with a single petal in $\diamondsuit$ along with three yellow petals in the complement, we discuss the alternating configuration.  First note that the placement of petals precludes the possibility of triggers in either $\eta_\diamondsuit$ or $\overline{\eta}_\diamondsuit$.  Further we note that here are the only instances where $\mathscr D$ may consist of multiple sets, where some transmission is actually needed.  Suppose then that $\mathscr D = \{\mathscr D_1, \mathscr D_2\}$, where $\mathscr D_1$ consists of a single petal and $\mathscr D_2$ the other two.  Then $\mathscr D_1$ is already connected and there is only one mixed mechanism to hook up $\mathscr D_2$, so the cost is $y+s$ which is the same as the blue transmission problem.  On the other hand, there may be several $\mathscr D_i's$ involved implying that a successful transmission of all of them requires all three yellow petals to be connected; in this case the only mechanism available is the pure yellow state in the iris.  Finally, for completeness, there is the case of a single $\mathscr D$ consisting of two of the petals while the third one is incidental.  This differs only formally from the $\mathscr D_1, \mathscr D_2$ case.   

For $\eta_\diamondsuit$ containing two yellow petals in the complement of $\diamondsuit$, there would be nothing to prove were it not for the advent of the triggering phenomena.  Indeed, all transmissions could only use a unique mixed hexagon and hence the probabilities would be just $y+s = b+s$.  However, unfortunately, the case of two yellow petals plus a conditioned yellow would often lead to triggering situations, boosting this probability to $\frac{1}{2}$.  Here we implement the appropriate dosage of no close encounter rules as before.  

The cases where $\diamondsuit$ consists of more than one petal are similar (or trivial).  At the level of conditional transmissions, given $\eta_\diamondsuit$, the full petal configuration,  these cases appear to be identical to the ones above with the r\^ole of the additional petals of $\diamondsuit$ played by petals of $\eta_\diamondsuit$ which happen to be the wrong color to aid transmission.  Notwithstanding, these problems are \emph{not} isomorphic, because of the advent of triggering in the comparisons of $\eta_\diamondsuit$ versus $\overline{\eta}_\diamondsuit$.  Nevertheless, the mechanisms exploited to handle to single petal problems do apply in the cases where $\diamondsuit$ has more than one petal.  Indeed, all that was needed to handle the single petal case was the explicit verification that the single petal of $\diamondsuit$ was in a position to influence the transmission.  Obviously, this will still be true in the multiple petal cases.  We see no merit in explicit calculations for these additional cases and therefore consider the proof to be completed.  
\qed

We now turn attention to cases where the conditioned hexagons include the iris.  Fortunately, the analogue of the above lemma, in its full generality, is certainly not necessary.  Indeed, it is important to realize that these exercises are tailored for situations where the conditioned hexagons in $\diamondsuit$ are, in fact, segments of paths.  These considerations drastically cut down the number of problems -- essentially to a single case, which we prove in the following:

\begin{lemma}\label{iris}
Let $\mathfrak F$ denote a flower and $\diamondsuit$ a specification of at least two petals and partial information about the iris with the property that a connection between two yellow petals of $\diamondsuit$ must be taking place through the iris.   Let $\mathscr D$ denote another set of petals on $\mathfrak F$ which is disjoint from $\diamondsuit$ and let $T_{\mathscr D, \diamondsuit}^B$ be defined as before.  Let $X^\odot_{\mathscr D, \diamondsuit}$ denote a $\{0, 1\}$ valued random variable and $T_{\mathscr D, \diamondsuit}^{Y*}$ the event that $\mathscr D$ is yellow connected such that: If $X^\odot_{\mathscr D, \diamondsuit} = 1$, usage of the iris is permitted, but, if $X^\odot_{\mathscr D, \diamondsuit} = 0$, usage of the iris is forbidden.  Then for $b \geq s$, there are joint laws such that
\begin{equation*}
\mu(T_{\mathscr D, \diamondsuit}^{B}) = \mu^*(T_{\mathscr D, \diamondsuit}^{Y*}),
\end{equation*}
where by abuse of notation from Lemma \ref{noiris}, $\mu^*$ denotes the appropriate joint distribution.  Similar results hold with the role of yellow and blue reversed and, in case $\diamondsuit$ has petals of both colors, $*$--probabilities for the $*$--transmissions of the two colors are equal.
\end{lemma}

\begin{remark}
In the non--trivial implementation of the above result, a clear interpretation of the above scenario is that the pure iris is shared by both ``paths''.  We adhere to this interpretation.
\end{remark}

\noindent
{\bf Proof: }  As in the Proof of Lemma \ref{noiris}, we will prove the analogue of Equation \ref{bayes}.  Due to the stipulation that $\diamondsuit$ must contain a yellow transmission through the iris, if the requisite transmission in $\diamondsuit$ is between diametrically opposed hexagons, the (conditional) blue transmission will occur automatically and there is basically nothing to prove.  
\begin{wrapfigure}{r}{0.48 \textwidth}
\vspace{-3mm}
\includegraphics[width=0.47 \textwidth]{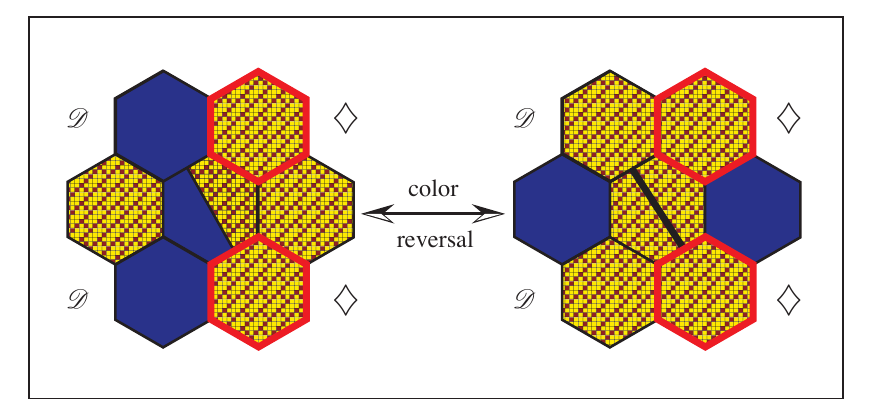}
\caption{\footnotesize{\label{illustration2}
All paths transmit through the iris.}}
\end{wrapfigure}
Indeed, the hexagons in $\diamondsuit$ plus the iris divide the remaining petals into two halves and, by micro--environment duality (c.f.~remark following Lemma \ref{conditioned}), there cannot be a blue connection between these two halves.  Evidently the only possible blue transmissions under consideration will be between adjacent petals.  In these cases we simply set $X^\odot_{\mathscr D, \diamondsuit} = 0$.  

Thus, the only non--trivial case is when there are two petals in $\diamondsuit$ separated by one unit with the appropriate mixed iris providing the required connection along with a pair of blue hexagons which are adjacent to this pair.  While perhaps not obvious in a verbal description, a look at Figure \ref{illustration2} shows that it is nevertheless true that the same mixed hexagon provides the requisite connection for $\mathscr D$.  Thus, in the presence of such an $\eta$, the conditional probability is
\begin{equation*}
\mu(T_{\mathscr D, \diamondsuit}^B \mid \eta) = \frac{s}{y+s}.
\end{equation*}
On the other hand, in $\overline\eta_\diamondsuit$, the only possibility for achieving the requisite yellow transmissions is when the iris is pure yellow which necessitates $X^\odot_{\mathscr D, \diamondsuit} = 1$.  Here we get
\begin{equation*}
\mu^*(T_{\mathscr D, \diamondsuit}^{Y*} \mid X^\odot_{\mathscr D, \diamondsuit} = 1, \overline{\eta}_\diamondsuit) = \frac{y}{y+s},
\end{equation*}
so if we adjust the conditional probability for $X^\odot_{\mathscr D, \diamondsuit} = 1$ to $s/y$, then the desired result is achieved.  
\qed

\begin{remark}\label{impunity}
It is important for later purposes to emphasize certain cases where the random variables do \emph{not} come into play:\\
\indent1. The random variables $X_{\mathscr D, \diamondsuit}$ are really defined conditional on the configuration $\eta_\diamondsuit$, i.e.~the entire petal configuration.  This has the following consequences: If the petal configuration is such that the required connection between say petal $x$ and $y$ has already occurred, then $X_{\mathscr D, \diamondsuit} \equiv 0$.  For later reference, we call such transmissions \emph{predetermined} transmissions.\\
\indent2. Our random variables are designed to punish or reward transmissions of the same color as the set being conditioned on and thereby level the playing field compared to transmissions of a different color.  In particular, if $\mathscr D$ is blue and $\diamondsuit$ is all yellow (or vice versa), then the random variables do not affect the transmission.
\end{remark}

We now recast the previous results in a form which is more pertinent for later use.

\begin{lemma}\label{conditioned}
Let $\Gamma_b$ be a blue path and let $\Gamma_y$ be a yellow path.  Let $x$ and $y$ be two points.  Then the probability of a $*$--transmission from $x$ to $y$ in the ``complement'' of $\Gamma_y$ and $\Gamma_b$ is the same in yellow as it is in blue.  Here, complementary $*$--transmission denotes, depending on the values of the auxiliary random variables and the relevant colors involved, the possibility of leeway provided for the sharing of hexagons and/or adherence to no close encounter rules, as discussed in Lemmas \ref{noiris} and \ref{iris}.
\end{lemma}
\noindent
{\bf Proof: }In light of the preceding two lemmas, all that is needed is an argument (involving inclusion--exclusion) along the lines used in the proof of Lemma \ref{unconditioned}.  We may follow the reasoning used therein mutatis mutandis.
\qed
\begin{remark}
We have made no stipulation about the path type of $\Gamma_b$ and $\Gamma_y$.  E.g.~self--avoiding, no close encounters, etc.  However, it turns out to be the case that if $\Gamma_y$ and/or $\Gamma_b$ were supposed to be self--avoiding in the strongest sense -- hexagon self--avoiding and no close encounters, then the presence of our additional transmissions do not change this property.  Indeed, the only mechanism for local changes in e.g.~the path $\Gamma_y$ is the transmutation of a mixed iris to a pure iris or vice versa.  Ostensibly, this could ``change'' the required use of a mixed iris in a path segment such as [3, 4, (mixed horizontal iris), 1] (in yellow) to a path where the use of 4 is redundant when the iris ``turns'' pure (c.f.~Remark \ref{self--avoiding?}).  However, under these and similar circumstances, the blue part of the iris, cannot, by micro--environment duality, be used to connect anything that cuts across the yellow path and the remaining petals of the flower, if used at all, will be automatically connected.  Hence, should the path $\Gamma_y$ have segments of this type, it will never be the case that the $*$--rules permit a change of the iris type.
\end{remark}

The following is of not immediate use but will be important later on.  We include the result here because the proof follows along the lines of what has preceded.
\begin{lemma}\label{full flower better}
Let $\mathfrak F$ be a flower and let $\diamondsuit$ and $\mathscr D$ be as in Definition \ref{diamond} and suppose that $a^2 \geq 2s^2$.  Then the probability of $\mathscr D$ being all of one color and connected in the same color conditioned on $\diamondsuit$ -- even with the $*$--rules enforced -- is no bigger than the same probability in the unconditioned case, e.g.
\begin{equation*}
\mu^*(T_{\mathscr D, \diamondsuit}^{B*}) \leq \mu(T_\mathscr D^B).
\end{equation*}
In particular, consider the event $\tilde{T}^{B*}_{\mathscr D, \diamondsuit}$ which is similar to $T^{B*}_{\mathscr D, \diamondsuit}$, but where the right to close encounters is never withheld.  Then
\begin{equation*}
\mu^*(\tilde{T}_{\mathscr D, \diamondsuit}^{B*}) \leq \mu(T_\mathscr D^B),
\end{equation*}
and similarly with $B$ replaced by $Y$.
\end{lemma}   
\noindent
{\bf Proof: }We discuss first the cases where $\diamondsuit$ does not include the iris.  We note that all situations where $\mathscr D$ consists of multiple sets do not actually involve the extra degrees of freedom provided by the random variable, so we in fact get the desired result immediately; usually as a strict inequality, i.e.~when the sites in $\diamondsuit$ are in a position to participate in the necessary connections.  Thus we may assume without loss of generality that $\mathscr D$ consists of two components which must be connected.   Let $\Box$ denote an alternative configuration to $\diamondsuit$ (on the same subset) and $\eta_{_\Box}$ the full configuration on all the petals.  Clearly it is enough to show 
\begin{equation}\label{good bad}
\sum_{_\Box, \eta_{_\Box}} \mu(\eta_{_\Box}) \mu(T_\mathscr D^B \mid \eta_{_\Box}) \geq \sum_{\eta_\diamondsuit}\mu(\eta_\diamondsuit) \mu^*(\tilde{T}_{\mathscr D, \diamondsuit}^{B*} \mid \eta_\diamondsuit).
\end{equation}
We divide into two cases, the first and more serious of which is when $\mathscr D$ contains next neighbor sites separated by a site which is not in $\mathscr D$.  However, if the site separating $\mathscr D$ is in $\diamondsuit$, the result is trivial:  Confining attention only to those configurations on ``the other side'' of $\mathscr D$, which, given the condition in $\diamondsuit$, would require a transmission, the difference between the left and right hand side is, at best, proportional to a $(b + 2s)$ for the transmissions with permissions, versus a  $\frac{1}{2}(b+s) + \frac{1}{2}$ times the same proportionality constant for the unconditioned case.  We may thus assume that the separating petal is not in $\diamondsuit$ and, obviously, since the terms in which it is blue contribute equally to the left and right side of Eq.~(\ref{good bad}), we may as well assume that this separating petal is yellow.  

We first consider the possibility that $\mathscr D$ contains more than just the two ``ports'' in question, i.e.~magnitude of $\mathscr D$ is bigger than or equal to 3.  If $|\mathscr D| \geq 4$ -- and there is no automatic transmission -- then the conditional transmissions will be $(a+2s)$ for both yellow and blue and therefore the $*$--rules do not even come into play.  Thus we have, for all $\Box$ configurations,
\begin{equation}\label{good bad petal}
\mu(T_{\mathscr D}^B \mid \eta_{_\Box}) \geq \mu^*(T^{B*}_{\mathscr D, \diamondsuit} \mid \eta_\diamondsuit),
\end{equation}
whenever $\eta_\diamondsuit = \eta_{_\Box}$ on the complement of the conditioned set.  Now, turning to cases where $|\mathscr D| = 3$, since $\mathscr D$ only has two components, the extra port must be contiguous to one of the other two.  The unconditioned case will be unity with probability $\frac{1}{4}$ (both petals not yet accounted for are blue), $\frac{1}{2}$ with probability $\frac{1}{4}$ (both yellow which leads to a triggering situation), and otherwise $(a+2s)$.  On the other hand, the conditional situation can at best get $(a+2s)$, which is smaller than the preceding combination.

We are down to the central cases we must consider: $|\mathscr D| =2$ and the two petals of $\mathscr D$ are separated by a single yellow petal.  The unconditioned case (under the above mentioned conditions) yields a grand total of:
\begin{equation}\label{magic number}
G_T = \frac{1}{8} \left[1 + 2\cdot\frac{1}{2} + 2 (b+s) + 3(b+2s)\right],
\end{equation}
where the various terms in the parenthesis are in obvious accord with each of the eight configurations.  Now we partition the remaining cases according to the size of $\diamondsuit$.  If $|\diamondsuit| = 3$, there is, in essence, nothing to prove unless there is a triggering situation.  Indeed, without triggers, the blue transmission probabilities (given $\eta_\diamondsuit$) and the yellow transmission probabilities (given $\overline{\eta}_\diamondsuit$) are identical and no $*$--rules would be implemented.  In the triggering situations, the best scenario for the conditional probability is $\frac{1}{2}$, which is easily exceeded by $G_T$.  

We are down to the case where $|\diamondsuit| = 2$.  If the two petals in $\diamondsuit$ are not contiguous, this, for all intents and purposes, reduces to the case where $|\diamondsuit| = 3$.  Indeed, the best scenario for the conditioned problem is a trigger, which leads to $\frac{1}{2} \leq G_T$.   For the remaining cases, we must treat separately the situations where both petals of $\diamondsuit$ are blue and when there is one blue and one yellow (we remind the reader that we need never consider the case where $\diamondsuit$ is entirely yellow in a blue transmission, c.f.~Remark \ref{impunity}).  In the case where $\diamondsuit$ is entirely blue, as far as the conditional transmission is concerned, when the unaccounted for petal is blue, there is no triggering and, at best, $(b + 2s)$; when the remaining petal is yellow, one gets $(a+s)$ (in both $\eta_\diamondsuit$ and $\overline{\eta}_\diamondsuit$ hence no rules are implemented).  Thus we are looking at equal admixtures of $(a+s)$ and $(a+2s)$, which is less than $G_T$.  Now, the final $|\diamondsuit| = 2$ situation: $\diamondsuit$ contains one yellow petal and one blue; we remind the reader that the two petals of $\diamondsuit$ are contiguous.  Summing over $\eta$, here we find equal admixtures of $\frac{1}{2}$ and $(a+2s)$ for the conditioned case; the second case is self--explanatory, the first case could directly be a trigger, or be an alternating configuration whose color reverse is a trigger.  In any case, a casual tally shows that $G_T \geq \frac{1}{4} + \frac{1}{2}(a+2s)$ and so we are done with $|\diamondsuit| = 2$.

We now turn to the consideration of $|\diamondsuit| = 1$, in which case this petal, wherever it may be located, is certainly blue.  If $\diamondsuit$ is contiguous with one of the ports, there will be a triggering scenario with probability $\frac{1}{4}$ (which is an enhancement over the color reverse) and to the rest of the configurations we assign $(a+2s)$.  However, we contend that $\frac{1}{4} \cdot \frac{1}{2} + \frac{3}{4} (a+2s)$ does not exceed $G_T$; this time, finally, due to the inequality $b \geq s$.  Finally, if $\diamondsuit$ is perched right between the two ports (on the ``big'' side), then in the non--triggering scenario, both unaccounted for petals of $\eta$ must be yellow, the color reverse of which does not even lead to triggering, therefore actually does worse than when the $\diamondsuit$ was contiguous with one of the ports.

The very last case to consider is where the two ports of $\mathscr D$ lie at opposite ends of the flower.  Borrowing from the previous next--nearest neighbor case, we may as well assume that these are the only petals of $\mathscr D$.  First the unconditioned probability ought to be computed.  As can be explicitly verified, along one route to connect $\mathscr D$ around the iris, the addition of either hexagon will already improve the probability to $y + 2s$; unfortunately, a single hexagon on the other side does nothing.  However, for this case of $\mathscr D$, by running the gamut of possibilities on the ``good'' and ``bad'' approaches, we still obtain
\begin{equation*}
\sum_{_\Box, \eta_{_\Box}}\mu(\eta_{_\Box}) \mu(T_\mathscr D^B \mid \eta_{_\Box}) = \frac{1}{4} + \frac{3}{4} \left(\frac{1}{4} + \frac{1}{4} (b + s) + \frac{1}{2} (b + 2s)\right).
\end{equation*} 
It is noticed that the term in parenthesis is in excess of $(b+2s)$, thus, even if $\diamondsuit$ is concentrated on one side of the ``transmission line'' -- which would produce a $\frac{1}{4}$ similar to the one in the above display; in every configuration in which there is no direct transmission, the conditional probability still does not exceed $(b+2s)$ and we are done.  

Finally, we discuss the circumstances where $\diamondsuit$ includes information about the irises.  While intricate arguments along the above lines are almost certainly possible, these problems are easily handled under the proviso $b^2 \geq 2s^2$ -- which is anyway implemented later for entirely different reasons.  Indeed, the only non--trivial cases, the ones discussed in the proof of Lemma \ref{iris}, are when the conditional transmissions are given by $s/(b+s)$.  On the other hand, given that $\mathscr D$ is blue, but in the absence of any other conditioning, a transmission always takes place with probability at least as big as $b+s$, which is greater than or equal to $s/(b+s)$, whenever $b^2 \geq 2s^2$.
\qed

\section{Convergence to Cardy--Carleson Functions}\label{conv_proof}
\subsection{\large{Introductory Remarks and More on Paths}}
Here we will introduce the functions, $u_{_N}^*$, $v_{_N}^*$ and $w_{_N}^*$, which are more or less the functions with which we will work.  Of course our ultimate theorem concerns the usual functions $u_{_N}$, $v_{_N}$ and $w_{_N}$ discussed in the introduction; but all of the mechanics, e.g.~Cauchy--Riemann relations, contour integration, etc., hold only for the former set.  We will be content with the knowledge that $|u_{_N} - u_{_N}^*| \rightarrow 0$ uniformly on compact sets disjoint from the boundary (which we do not ultimately prove till the appendix) and similarly for the $v$'s and $w$'s.  For the purpose of what is to follow, let us introduce some concise notation.  

\begin{nota}
Let $\mathcal D \subset \mathbb C$ denote a finite, open, simply connected domain with piecewise smooth boundary, which we will regard as having a diameter of order unity.  The boundary of $\mathcal D$ is exhausted by three disjoint (except possibly for end points) connected sets, which we denote by $\mathscr A$, $\mathscr B$ and $\mathscr C$, in counterclockwise order.  We tile $\mathcal D$, including the boundary, with hexagons of scale $N^{-1}$, and we will freely use the notation $\mathscr A$, $\mathscr B$ and $\mathscr C$ to denote the boundary hexagons corresponding to these three boundary pieces.  While there may be some ambiguity as to which boundary piece a few hexagons belong to, we do not dwell on these details; it is sufficient that some choice be made which keeps these sets connected.  The resulting subset of the hexagon lattice we will denote by $\Lambda^{(N)}$ and we will place a floral arrangement $\Lambda^{(N)}_{\mathfrak F_N}$ inside $\Lambda^{(N)}$ in accord with the conventions discussed in Section \ref{def}.  Since all of the actual labor will take place at finite $N$, we will, whenever possible, treat the hexagons as separated by unit distances and simply regard $N$ as a large parameter.  In particular, we use the notation $z$ to locate vertices of the hexagon lattice; most of our $z$'s will be of order $N$.      
\end{nota}

As was the case in \cite{Smir}, the functions are defined on the vertices of the hexagons and smoothly extended if technically necessary.  Let us focus on the $u$'s since the same considerations hold for $v$'s and $w$'s.  We start with a definition of the standard $u_{_N}(z)$ in blue, which is the probability of the following event: There is a blue path from $\mathscr A$ to $\mathscr B$, separating $z$ from $\mathscr C$.  To be definitive, the path must be self--avoiding but with close encounters permitted; as will be demonstrated in the appendix, such matters are inconsequential in the large $N$ limit.  We define $\mathscrb U_{_N}(z)$ to be the indicator function of the event just described.  We will not be notationally specific as to whether we are talking about a blue path or a yellow path for this event; in any case, we define $u_{_N}(z) = \mathbb E [\mathscrb U_{_N}(z)]$.  

The function $u_{_N}^*(z)$ is analogous to $u_{_N}(z)$ in that both concern the probability of a path from $\mathscr A$ to $\mathscr B$ that separates $z$ from $\mathscr C$.  However, first we should emphasize that $u_{_N}^*$ pertains to a probability on our enlarged space and second, there are the seemingly modest differences which become very important in the (unlikely) event that the path comes close to $z$.  In fact, at the finest level of distinction, our functions will be the expectations of random variables rather than the probabilities of events.  While there are again two versions of our functions, one for yellow and one for blue, for ease of notation we will still omit specific reference to the color, and, for the sake of definitiveness, unless otherwise specified we will be talking about the blue version of these objects.

We turn to the definition of the object $\mathscrb U_{_N}^*(z)$, a random variable, which defines $u_{_N}^*(z)$.  In most cases, $\mathscrb U_{_N}^*(z)$ is in fact the indicator of an event and $u_{_N}^*(z)$ the corresponding probability; we will proceed with this language and later highlight the configurations in which the random variable takes on a value other than zero or one.  First and foremost, $\mathscrb U_{_N}^*(z)$ indicates an event on $\Omega_N \hspace{-1mm}\times \mathbb D^K$, where $\Omega_N$ is the set of percolation configurations in $\Lambda_{\mathfrak F_N}^{(N)}$, $K$ is the number of flowers in $\Lambda^{(N)}_{\mathfrak F_N}$ and $\mathbb D$ is the space corresponding to the range of the random variables $X_{\mathscr D, \diamondsuit}$ and $X^\odot_{\mathscr D, \diamondsuit}$.  In order for a configuration to satisfy the criterion of $\mathscrb U_{_N}^*(z)$, it is first necessary that the hexagons contain a blue path connecting $\mathscr A$ and $\mathscr B$ separating $z$ from $\mathscr C$.  As of yet we make no specifications concerning the type of the path -- it may contain close encounters and it may contain shared hexagons.  Note that a path can be ``contracted'', i.e.~by cutting out loops till it is a self--avoiding, non--self--touching path.  The resulting path still connects $\mathscr A$ to $\mathscr B$ and, if it still separates $z$ from $\mathscr C$ (which need \emph{not} be the case) then, as we shall see, the event $u_{_N}^* (z)$ is automatically satisfied regardless of the auxiliary variables.  It is in the grey zone between the extremes of $\{$no separating path exists$\}$ and $\{$a separating path exists which enjoys strict self--avoidance$\}$ where the random variables $X_{\mathscr D, \diamondsuit}$ and $X^\odot_{\mathscr D,\diamondsuit}$ really come into play.  

In order to be concrete, we will simply give a prescription which shows whether a particular path $(h_1, \dots, h_M)$ of blue and mixed hexagons in a configuration $\omega$ satisfies, depending on the values of the $X_{\mathscr D, \diamondsuit} ~\& ~X_{\mathscr D, \diamondsuit}^\odot$'s, the event $\mathscrb U^*_{_N}(z)$.  First and foremost, the underlying segments which form a ``skeleton'' for the blue path must constitute an actual self--avoiding path from $\mathscr A$ to $\mathscr B$ which separates $z$ from $\mathscr C$.  Thus, the hexagons have been ordered in such a way that the skeleton does not cross itself.  Second, in the region complementary to flowers (if any), the path must obey the ``conventional'' rules, i.e.~no sharing of hexagons permitted, self--touching allowed.  We now turn to the delicate discussion of what takes place within the flowers.  The best prescription is to follow the path sequentially: by and large, the first pass of the path through any flower is ``free''.  If the flower is never revisited, it need not be considered again, but, in case the path returns to the flower, the initial portion of the flower which had been used defines, temporarily, the set $\diamondsuit$.  The value of $X_{\mathscr D, \diamondsuit}$ for all possible $\mathscr D$'s is now ascertained.  When the path revisits the flower, with the intention to share a hexagon of $\diamondsuit$, or, encounter a hexagon of $\diamondsuit$, it must receive ``permission'' from the appropriate $X_{\mathscr D, \diamondsuit}$ and/or $X_{\mathscr D, \diamondsuit}^\odot$.  If success is achieved at this level, the new $\diamondsuit$ is reset by adjoining to the old $\diamondsuit$ the petals that had been used in the second visit; all of this in case of a possible third visit, etc.  Failure on any pass through any flower renders that particular path useless for achieving the event.  Notwithstanding, \emph{all} candidate paths must be checked; if no path of $\omega$ satisfies the geometric criterion with these permissions, then the event $\mathscrb U_{_N}^*(z)$ does not occur.  If at least one path satisfies all of the above criteria, then $\mathscrb U_{_N}^*(z)$ is declared to have occurred.  The event $\mathscrb U_{_N}^*(z)$ has been defined; corresponding definitions hold for $\mathscrb V_{_N}^*(z)$ and $\mathscrb W_{_N}^*(z)$.

The exceptional situations occur when $z$ is a vertex of an iris hexagon \emph{and} the path under consideration ostensibly goes through the iris.  It is worthwhile, referring to the previous discussion, to assign a value to each path, namely zero or one, and then define $\mathscrb U_{_N}^*(z)$ to be the maximum over all paths of the path value.  We will continue this perspective.  Let us now describe the circumstances under which the path value will be set to $\frac{1}{2}$: First, as alluded to, $z$ itself must be the vertex of an iris hexagon; second, the iris must be in a mixed state; and finally, the path under consideration would lead to a value of one \emph{if} the iris had been pure blue (and of course zero had the iris been pure yellow).  Notice that depending on the particulars of the mixed state and the path, the path value can be $\frac{1}{2}$ even when the requisite blue transmit has not literally occurred.  Under these circumstances, $\mathscrb U_{_N}^*(z)$ \emph{may} take on the value $\frac{1}{2}$.  Of course, it should be emphasized that if an alternative path exists which does not use the iris and does satisfy all the requisite permissions, then $\mathscrb U_{_N}^*(z)$ will be one.  Thus it is only the configurations in which the iris attached to $z$ is \emph{pivotal} for the relevant event that $\mathscrb U_{_N}^*(z)$ can be $\frac{1}{2}$.  These are, as is well known from \cite{Smir}, exactly the configurations contributing to the derivatives of the relevant functions.

As is seen from the above descriptions, it is indeed the case that anytime a self--avoiding, non--self--touching path of the right color separates $z$ from $\mathscr C$, $\mathscrb U^*_{_N}(z) = 1$, simply because no permissions are ever required.  Thus without the advent of sharings, etc.~no such paraphernalia would be necessary and we might just as well focus on the \emph{reduced} path.  However, it is crucial to our analysis that certain paths loop around in order to ``capture'' $z$.  Nevertheless, the existence of certain self--avoiding, non--self--touching paths is important for conditioning/partitioning purposes.  In this vein, one might envision that a path with permissions which nevertheless contain such loops may be partially reduced in this fashion, i.e.~the journey ``towards'' $z$ indeed has this property, with all the auxiliaries occurring in the later portion of the path.  That such a rearrangement is possible is the subject of the next lemma.

\begin{defn}
Consider a blue transmit in the configuration $\omega$ which satisfies the (geometric) requirements of the event that $\mathscrb U^*_{_N}(z) \neq 0$.  If this path cannot be reduced to a self--avoiding, non--self--touching path then it has loops which are essential for the fulfillment of this event.  We define the \emph{lasso} points of this path as follows: The last lasso point is a shared hexagon or a close encounter pair which is part of a relatively simple closed loop of the path with $z$ in its interior.  The next to last lasso point (if any) enjoys a similar definition, save that the loop in question passes through the last lasso point.  Similarly for the earlier lasso points.  
\end{defn}

\begin{lemma}\label{loop erasure}
Suppose that $(\omega, X)$ is a configuration such that $\mathscrb U_{_N}^*(z) > 0$.  Then, in $\omega$ there is a path fulfilling the requirements of $\mathscrb U_{_N}^*(z) > 0$ (i.e.~connects $\mathscr A$ to $\mathscr B$ and separates $z$ from $\mathscr C$) with the property that in the part of the path from $\mathscr A$ to the last lasso point necessary for the capture of $z$, the only points of sharing or pairs of close encounters are those which are essential for the particular path to fulfill the criterion $\mathscrb U^*_{_N}(z) > 0$.  In particular this portion of the path may be regarded as having no sharings and no close encounters with itself.
\end{lemma}

\begin{remark}
We remark that while the above appears to be geometrically obvious -- just cut out the necessary loops -- what is at issue is that the rearranged path still has some close encounters/shared hexagons with the later portion of the path.  Thus it is not \emph{a priori} clear that the new path, with the new $\mathscr D$'s, will still have the requisite permissions.  In point of fact, the stronger statement that the \emph{full} path can be reduced to one in which all the shared hexagons and close encounters remaining are essential for the capture of $z$ turns out to be false, as the following example shows.
\end{remark}

\begin{ex}
We consider a situation -- destined for a \emph{yellow} capture of $z$ -- in which the initial incoming line to the flower is at petal 3 whereupon the path leaves the flower immediately and, after capturing $z$, returns to petal number 6.  It then leaves again and reenters at petal 5 (thereby making a redundant loop), undergoes a diametric transmission through the iris to petal number 2 and leaves for the last time.  Notice that petals 1 and 4 have not been specified, but let us assume that they are both blue.  The initial condition for transmission -- before the reduction -- is that petals 6 and 3 are conditioned on; however, after the reduction, we regard the reentrance -- after capture -- at petal 6 to be a fresh transmit to 2, where petal number 5 happens to be yellow.  Thus, in the reduced version of the transmission problem, $\diamondsuit$ consists solely of petal 3.  The reader can check that for this transmission situation, both the $\beta$ and $\gamma$ ($60^\circ$ and $120^\circ$) mixed hexagons will provide the requisite transmission, so the overall un--starred transmission probability would be $(a+2s)$.  On the other hand, the color reverse of this scenario (keeping the singleton in $\diamondsuit$ fixed at yellow) represents a trigger situation, so, indeed, the reduced transmission will require permissions for a close encounter with the conditioned petal at 3.  
\end{ex}

\noindent
{\bf Proof: }Any reduction of the requisite type that takes place on the complement of flowers may, obviously, be performed without discussion.  We are therefore, without loss of generality, down to the consideration of paths where all loop and lasso points take place within flowers.  Now suppose a flower only contains loop points whose removal does not affect the separation event.  Then, as discussed previously, we claim that the required reduction may also be performed with impunity.  (To recapitulate, if the reduction within the flower can be performed which then renders the path segment going through a flower as self--avoiding and non--self--touching, then, in the new path within the associated flower, no random variables need to be consulted since no permissions are actually required.)  

We will consider a flower $\mathfrak F$ which contains a generic lasso point of the separation event, and let $\Gamma$ denote the (unreduced) path which actually satisfies the event.  More precisely, $\Gamma$ will enter the flower at some petal $e_0$ and, after some meandering (possibly leaving the flower to make redundant loops) must leave the flower at some petal $c$ to capture $z$; the petal $c$ is defined by the condition that it is the last petal of $\mathfrak F$ that $\Gamma$ visits before capturing $z$, i.e.~the next time $\Gamma$ visits $\mathfrak F$ it will have generated a loop with $z$ in its interior.  We therefore need to show that the part of $\Gamma$ between $e_0$ and $c$ -- which we denote by $\Gamma_{\mathfrak F}$ -- can be made strongly self--avoiding.  Denoting the reduced path by $\hat{\Gamma}_{\mathfrak F}$, we need to guarantee that $\hat{\Gamma}_{\mathfrak F}$ is actually a legitimate path.  The cases we have to treat are the ones in which there are one or more loop points in $\mathfrak F \cap \Gamma_{\mathfrak F}$ and for the event to be accomplished, we must make another essential non--predetermined transmission through the flower before we get to $c$ (c.f.~Remark \ref{impunity}).  We reiterate that these cases are dangerous because after the removal of the loop, the corresponding $\diamondsuit$ we condition on (to make the transmission) may change so it is not \emph{a priori} clear that the random variable will still ``allow'' the required transmission to happen.  Nevertheless, we have a fairly limited situation and we are able to ensure that the necessary transmission does indeed happen after the reduction.

We consider $\hat{\Gamma}_{\mathfrak F}$ and make the following definitions for convenience.  First, within the petal, the three hexagons -- including the iris -- which form the non--predetermined core of the transmission will be call the \emph{transmission line}; we also denote the first petal in the path ordering of the transmission the \emph{port} and the last petal in the transmission the \emph{terminus}.  

We start by focusing our attention on the case where no hexagon was shared.  Then we have two cases corresponding to whether the port and the terminus are diametrically opposed or next nearest neighbors.  We observe that $e_0$ cannot be next to the port or the terminus, because in the former case it would be directly connected to the port, hence in $\hat{\Gamma}_{\mathfrak F}$ there is no conditioning to be spoken of so the corresponding random variable is identically $\textsc o$.  In the latter case, since the capture of $z$ is purported to take place after the transmission, said transmission is not actually necessary to get to the terminus.  The situation is even more trivial if the port or the terminus is equal to $e_0$.  This implies that we are done with the case where the port and the terminus are diametrically opposed.  The second geometry follows similarly: $e_0$ cannot be on the small side of the transmission line and, indeed, can only occupy the mid petal of the large side of the transmission line.  Now if $\Gamma_{\mathfrak F}$ used the petal between $e_0$ and the port at all, then we are automatically done because then in $\hat{\Gamma}_{\mathfrak F}$, we have an unconditioned transmission between the port, $e_0$, the petal between them and the terminus.  On the other hand, if $\Gamma_{\mathfrak F}$ did not use the petal between $e_0$ and the port, then either $\Gamma_{\mathfrak F} = \hat{\Gamma}_{\mathfrak F}$ (the iris exhibits exactly the mixed configuration connecting the port to the terminus -- necessitating an eventual departure before $e_0$ connects to the port) or the iris was pure and we have a unconditioned situation where $e_0$ is connected directly to the terminus through the iris.

We now turn attention to the cases where there is sharing.  Our first claim is that under any circumstances of multiple passes through the same flower, there cannot be more than one instance of sharing.  Indeed, suppose there were two instances of sharing, then a rudimentary countings of any double sharing scenario demonstrates that at least five petals must be involved.  Thus in the first pass through the flower which requires sharing, the minimal situation is one conditioned hexagon in $\diamondsuit$ and four petals already blue in $\eta_\diamondsuit$.  These are precisely the circumstances which were discussed at the beginning of the proof of Lemma \ref{noiris} and thus no sharing is permitted on this first attempt to share.  On the other hand, if two petals are conditioned on before the first sharing -- so that now all remaining petals are blue -- any scenario either leads to  probability one transmission situations or, at worst, the scenario where there is just one mixed iris which fails to allow the desired transmission, with the same being true for the color reverse, hence no sharing again.  If three or less sites are left over after the first pass, there are not enough sites left for two or more passes involving transmission through the iris.  

Given the claim that there will be only one sharing we can divide into the cases where petals are being shared and where the iris itself is being shared.  The later has severe constraints, since the two transmissions must be side by side (c.f.~the proof of Lemma \ref{iris}).  In a straightforward rendition where the two transmissions are anti--parallel, both transmissions are redundant in the ultimate use of the flower, since the last entrance before the transmissions and the first exit after the transmissions are neighbors.  The less straightforward renditions of parallel transmissions appear to be a topological impossibility given what the rest of the path is purported to do.  Nevertheless, the shortened path now has a diametric transmission with two unconditioned blue petals, one on each side of the transmission axis, and at least one more (unconditioned) petal known to be blue due to a future visit of the flower after the capture of $z$.  

Finally, let us consider, in general terms, the (single sharing) situations where petals are shared during transmission.  Here we will only make intermittent reference to whether we are discussing the path before or after the reduction.  First, the flower must be visited and departed from without transmission, perhaps multiple times -- in order that there would be something to condition on when transmission finally occurs.  We claim that for such a transmission, we need only discuss cases where the port and terminus are both separated from the conditioned set by at least one spacing.  If not, the path under consideration is evidently the before path and the after path can get directly to the port or terminus thereby implying an unconditioned transmission or an unnecessary transmission, respectively.  Now, for the remaining cases, it is clear that the conditioned set is but a single petal.  Indeed, the geometry of conditioned site, port and terminus, is the previously discussed alternating pattern.  We claim that one of the three petals which are as of yet unaccounted for must be blue since, as the reader will recollect, the path is destined to return after the capture of $z$.  We now discuss two cases.  First the iris is pure blue, in which case, once again, we are evidently referring to the path before reduction since this \emph{can} be reduced.  However, the reduced path would then have an unconditioned transmission from the conditioned site to the terminus, which requires no permissions from random variables.  Otherwise, a more serious sort of transmission is taking place, evidently through a mixed iris.  Under these conditions, according to the conditional distributions, there will be no sharing permitted unless, possibly, the remaining two unaccounted for sites are both yellow.  The mixed type of the iris is now uniquely specified, and, due to the alternating geometry, does \emph{not} allow the direct transmission between the conditioned petal and the terminus.  But now, in as far as these visits to the flower are concerned, the path is in fact self--avoiding and non--self--touching.  Due to the constraints which led to the circumstances, there is/was no possibility for reduction, i.e.~it appears that we are looking at both the before and the after path with no need for analysis.       
\qed

\subsection{Statement and Proof of Cauchy--Riemann Relations}
In this section we will establish Cauchy--Riemann relations for the triple of functions under consideration.  As was the case in \cite{Smir}, these are not exactly Cauchy--Riemann \emph{equations}, but equations of a Cauchy--Riemann type between positive and negative ``pieces'' of the derivative, which admit a probabilistic interpretation.  Notwithstanding the absence of Cauchy--Riemann \emph{equations}, these Cauchy--Riemann relations are sufficient to exhibit Green's Theorem type cancellations in the evaluation of the appropriate discrete contour integrals.

\begin{defn}
Let $\hat a = i, \hat b = \tau i, \hat c = \tau^2 i$ denote three of the six lattice directions on the hexagonal lattice, where $\tau$ $=$ exp$(\frac{2\pi i}{3})$.  For a function $f(z)$ defined on the vertices of the hexagonal lattice and $\eta \in \{\pm\hat{a}, \pm\hat{b}, \pm\hat{c}\}$, as appropriate, we define the directional derivative in the usual fashion:
\begin{equation*}
D_\eta f (z) = f(z+\eta) - f(z). 
\end{equation*}
\end{defn}

Let $\mathscrb U_{_N}^{\mbox{\tiny{B}}*}(z)$, $\mathscrb V_{_N}^{\mbox{\tiny{B}}*}(z)$ and $\mathscrb W_{_N}^{\mbox{\tiny{B}}*}(z)$ denote the blue versions of the random variables described in the previous subsection and $\mathscrb U_{_N}^{\mbox{\tiny{Y}}*}(z)$, $\mathscrb V_{_N}^{\mbox{\tiny{Y}}*}(z)$ and $\mathscrb W_{_N}^{\mbox{\tiny{Y}}*}(z)$ their yellow counterparts.  We denote by $u_{_N}^*(z)$, $v_{_N}^*(z)$ and $w_{_N}^*(z)$ the expectation of the color neutral averages, e.g.
\begin{equation*}
u_{_N}^*(z) = \frac{1}{2}\mathbb E[\mathscrb U_{_N}^{\mbox{\tiny{B}}*}(z) + \mathscrb U_{_N}^{\mbox{\tiny{Y}}*}(z)],
\end{equation*}
and similarly for $v^*$ and $w^*$.  The \emph{Cauchy--Riemann pieces} are the quantities
\begin{equation*}
[u_{_N}^*]^+_\eta = [u_{_N}^*(z)]^+_\eta = \mathbb{E} \left[\left[\left(\mathscrb U_{_N}^{\mbox{\tiny{B}}*}(z+\eta) + \mathscrb U_{_N}^{\mbox{\tiny{Y}}*}(z+\eta)\right) - \left(\mathscrb U_{_N}^{\mbox{\tiny{B}}*}(z) + \mathscrb U_{_N}^{\mbox{\tiny{Y}}*}(z)\right)\right]^+\right]
\end{equation*}
\begin{equation*}
[u_{_N}^*]^-_\eta = [u_{_N}^*(z)]^-_\eta = \mathbb{E} \left[\left[\left(\mathscrb U_{_N}^{\mbox{\tiny{B}}*}(z+\eta) + \mathscrb U_{_N}^{\mbox{\tiny{Y}}*}(z+\eta)\right) - \left(\mathscrb U_{_N}^{\mbox{\tiny{B}}*}(z) + \mathscrb U_{_N}^{\mbox{\tiny{Y}}*}(z)\right)\right]^-\right],
\end{equation*}
where $(~~~)^\pm$ means positive/negative part and, typically, we will suppress the $z$ dependence.  Similar definitions hold for the quantities $[v_{_N}^*]^\pm_\eta$ and $[w_{_N}^*]^\pm_\eta$.  Of course we have $D_\eta u_{_N}^*(z) = [u_{_N}^*]^+_\eta - [u_{_N}^*]^-_\eta$, and similarly for $v^*_{_N}$ and $w^*_{_N}$.  We note that, in reference to the above display, there could be a distinction between ``the positive parts of the sum'' and ``the sum of the positive parts''.  However, as we shall see, in any configuration where, e.g., $(\mathscrb U_{_N}^{\mbox{\tiny{B}}*}(z+\eta) - \mathscrb U_{_N}^{\mbox{\tiny{B}}*}(z)) > 0$, the corresponding yellow term automatically vanishes.  A statement of the Cauchy--Riemann relations is as follows:

\begin{lemma}\label{CR}
Consider the Cauchy--Riemann pieces as described above.  Then, between $u$ and $v$, these objects satisfy six Cauchy--Riemann relations, the first three of which are:
\begin{equation*}
[u_{_N}^*]^+_{\hat{a}} = [v_{_N}^*]^+_{\hat{b}};~~[u_{_N}^*]^+_{\hat{b}} = [v_{_N}^*]^+_{\hat{c}};~~[u_{_N}^*]^+_{\hat{c}} = [v_{_N}^*]^+_{\hat{a}}
\end{equation*}
for site $z$ which emanate the edges $\hat a$, $\hat b$ and $\hat c$.  For sites emanating the edges $-\hat a$, $-\hat b$ and $-\hat c$, we have:
\begin{equation*}
[u_{_N}^*]^+_{-\hat{a}} = [v_{_N}^*]^+_{-\hat{b}};
~~[u_{_N}^*]^+_{-\hat{b}} = [v_{_N}^*]^+_{-\hat{c}};
~~[u_{_N}^*]^+_{-\hat{c}} = [v_{_N}^*]^+_{-\hat{a}}.
\end{equation*}
We note that
\begin{equation*}
[u_{_N}^*(z)]^-_{\hat a} = [u_{_N}^*(z+\hat a)]_{-\hat a}^+,
\end{equation*}
and similarly for $\hat b$ and $\hat c$, so the above implies all the necessary relationships for the negative pieces.  There are six corresponding equations between the derivative pieces of the $v$ and $w$ functions (which implies an additional six relations between the derivative pieces of the $w$ and $u$ functions).
\end{lemma}

We will prove separately the cases for sites which are and are not vertices of irises.  
\medskip

\noindent
{\bf Proof (non--iris sites): }If neither $z$ nor its neighbor is the vertex of any iris, the preliminary step of the proof is identical to that in \cite{Smir}.  Explicitly, let us consider the case of $[u_{_N}^*]^+_{\hat a}$.  Since no mixed hexagon is involved, both the blue and yellow versions correspond to the \emph{event} that the separating path goes ``below'' $z+\hat a$ but does not go ``below'' $z$.  Hence, focusing attention on the function $u_{_N}^{\mbox{\tiny{B}}*}(z)$, it is the case that the hexagons surrounding the edge $<\hspace{-1mm}z, z+\hat a\hspace{-1mm}>$ are both blue, while the one directly ``below'' $z$ is yellow; we will informally refer to these three hexagons as a \emph{triad}.  Note that by this criterion (among several others) no configuration will contribute to both the positive part of the blue piece \emph{and} the positive part of the yellow piece.  Returning attention to the blue case, the yellow hexagon in the triad is the terminus of a yellow path connecting to the domain boundary $\mathscr C$; for all intents and purposes, this path may be regarded as self--avoiding and non--self--touching.  As for the former pair, we may regard these as neighbors in a legitimate blue path which starts at $\mathscr A$, goes through these two from right to left and ends at $\mathscr B$.  By Lemma \ref{loop erasure} we may, without loss of generality, regard the first portion of the path, namely that which connects $\mathscr A$ to the hexagon on the right of $<\hspace{-1mm}z, z+\hat a\hspace{-1mm}>$, as self--avoiding and non--self--touching.  From the perspective of the remaining blue hexagon, what is required is therefore a conditional transmission -- with all rules enforced -- starting at this point and ending at $\mathscr B$.  (Note also that this path may have collisions, i.e.~sharings of mixed hexagons with the yellow path, but as for its \emph{interaction} with the yellow path, of course, no permissions are required.)  We will replace this transmission with the same sort of transmission in yellow, after some partitioning.       

We claim, according to standard arguments, that given the existence of a self--avoiding, non--self--touching blue path from $\mathscr A$ to the right hexagon of the triad \emph{and} a yellow path from the bottom hexagon of the triad to $\mathscr C$ -- i.e.~some sort of path from $\mathscr A$ to $\mathscr C$ -- there is a ``lowest'' such path.  We remark that all of the pure irises involved in these paths are of the obvious requisite type, and sometimes the mixed hexagons will be completely specified by the local geometry of the path, while in other cases it may be ambiguous.  With the latter consideration, we are therefore in fact conditioning on a path event rather than an actual path.  It is, however, clear that details of the configuration outside the path will in fact dictate the nature of certain irises.  In particular, one can envision a scenario where had the iris been pure yellow, due to some local deviation, an alternative path would have indeed been lower; therefore this mixed iris must be of a particular type.  Ostensibly we will run into a dual aspect of this situation: under certain circumstances, the newly formed yellow path will be allowed to share an iris, thereby (effectively) turning a mixed hexagon into a pure hexagon.  In light of the previous consideration, while the transmission may be successful, this switching could disrupt the conditioning.  However, as is not hard to see, these scenarios cannot come to pass.  Indeed, we claim that if changing the status of an iris from mixed to pure produces a lower path, it must be the case that the blue portion of the iris is, in fact, already in the region below what was previously the lowest path.  To demonstrate this, one only need to appeal to the skeleton structure of the underlying path: if it is  possible to lower the path by switching the blue half into a pure yellow, the closure of the symmetric difference of the lowest possible skeleton of the old path and the lowest possible skeleton of the new path forms a closed loop with the blue half of the hexagon in its interior, which concludes the demonstration.  We may therefore conclude that any iris involved in the yellow portion of the lowest yellow--blue path is either frozen into a particular mixed state -- with the blue portion of the hexagon inside the conditioned region and therefore inaccessible for sharing -- or is of a nature such that transforming the iris into a pure yellow does not render a change in the the condition of the lowest path.

It is now clear that modulo some necessities regarding triggering possibilities of the flowers which have been traversed by these paths, the region above this ``lowest'' blue--yellow path is entirely unconditioned.  We are therefore in a position to apply Lemma \ref{conditioned} (which automatically accounts for the triggering scenarios) to conclude that the conditional probabilities associated with the blue version of $[u_{_N}^*]^+_{\hat a}$ and the yellow version of $[v_{_N}^*]^+_{\hat b}$ are identical.  Running the same argument for the yellow version of the function $[u_{_N}^*]^+_{\hat a}$ and the blue version of the function $[v_{_N}^*]^+_{\hat b}$, we conclude $[u_{_N}^*]^+_{\hat{a}} = [v_{_N}^*]^+_{\hat{b}}$.  The other 11 relationships, for the non--iris sites, follow from an identical argument.

\medskip
\noindent
{\bf Proof (iris sites): }For convenience, we will start with the $\hat a$ derivative  of $u_{_N}^*(z)$, assuming the iris is located directly to the right of $<\hspace{-1mm}z, z+\hat a\hspace{-1mm}>$.  We first note that in those configurations where the iris happens to be pure, the argument is identical to the non--iris site case.  So we will focus attention on configurations contributing to $[u_{_N}^*(z)]^+_{\hat a}$ in which this iris is of a mixed type.  Our first case will be to compare the positive part of the $\hat a$ derivative of $u_{_N}^*$ to the positive part of the $\hat c$ derivative of $w_{_N}^*$.  Notice that in this case -- as opposed to an $\hat a$ versus $\hat b$ comparison -- the edges $<\hspace{-1 mm}z, z+\hat a\hspace{-1mm}>$ and $<\hspace{-1mm}z, z+\hat c\hspace{-1mm}>$ are both boundary edges of the iris and hence the situation before and after the switch will be more or less equivalent.  We start by considering configurations for which $\mathscrb U_{_N}^*(z+\hat a)=1/2$ while $\mathscrb U_{_N}^*(z) = 0$.  Aside from the mixed nature of the iris, we claim this is exactly the same as the pure iris case.  Indeed, the inferred value of $\mathscrb U_{_N}^*(z+\hat a)$, were this iris blue, is supposed to be one, while the inferred value of $\mathscrb U_{_N}^*(z)$ is still zero, meaning that the hexagon to the left of the $<\hspace{-1mm}z, z+\hat a\hspace{-1mm}>$ bond is indeed blue (and connected to $\mathscr B$), and similarly the hexagon below $z$ is yellow, etc.  Now, it is only necessary to observe that changing the iris to yellow destroys the event of a separating path ``below'' $z+ \hat a$, which is indeed seen to be the case.  For this portion of the proof, we will actually do a double switch: first changing the blue path from the left hexagon to $\mathscr B$ to yellow and then replacing the yellow path which connects to $\mathscr C$ with a blue rendition.  The former is identical to the argument of the pure case modulo that we must envision the mixed hexagon as a pure blue in order to perform the conditioning partition.  Having accomplished the first switch, we claim that the second switch is identical -- with the same proviso concerning the mixed hexagon and, of course, a repartitioning of the configurations according to the ordering of the new yellow--blue path connecting $\mathscr B$ to $\mathscr A$.  When the double procedure has been achieved, we are, manifestly, in a configuration where the blue version of $\mathscrb W_{_N}^*(z+\hat c)$ evaluates to $1/2$ while, still, the corresponding version of $\mathscrb W_{_N}^*(z)$ is zero.  Since by a rotation of the arguments at the beginning of this paragraph, these are the only such configurations contributing to (the positive part of) $\mathscrb W_{_N}^*(z+\hat c) - \mathscrb W_{_N}^*(z)$ (in blue), and hence we have a bijection between the configurations contributing to the positive part of the $\mathscrb U^*_N$ difference (in blue) and the positive part of the $\mathscrb W^*_N$ difference (in blue).   

Finally, starting from the same initial setup, we now compare the $\hat{a}$ derivative of $u_{_N}^*(z)$ with the $\hat b$ derivative in $v^*$.  As alluded to above, this case is essentially different because the site at $z+\hat b$ is actually surrounded by pure hexagons.  Proceeding in the \emph{forward} direction, we follow the steps of the pure case: that is to say, we replace the blue path emanating from the hexagon to the left of $<\hspace{-1mm}z, z+\hat a\hspace{-1mm}>$ with a yellow transmission.  Let us investigate the consequences.  It is clear that $\mathscrb V_{_N}^*(z+\hat c)$ indeed equals one (regardless of the iris configuration) and now we claim that $\mathscrb V_{_N}^*(z) = 1/2$.  Indeed, in light of the two hexagons below and to the left of $z$, through which a yellow path connects $\mathscr B$ to $\mathscr C$, it is clear that were the iris yellow, the yellow version of $\mathscrb V_{_N}^*(z)$ would be one; however, the blue path which connects the outside of this iris to $\mathscr A$ indicates that were the iris to be blue, no yellow path would separate $z$ from $\mathscr A$.  We are finished with the forward direction.  The last thing to be checked is that the map we just described onto, which amounts to the statement that in any configuration where $\mathscrb V_{_N}^*(z+\hat b) = 1$, while $\mathscrb V_{_N}^*(z) = 0$ (in yellow) is of the above described form.  But here the argument runs a very close parallel to the considerations at the beginning of the previous paragraph: By assumption, the iris is in a mixed state, but even if the iris were blue, there must be a yellow separating path to the right of $z + \hat c$, and this forces the two pure hexagons of the appropriate triad to be yellow.  Envisioning the iris as yellow places a path to the right of $z$; however, when this iris is blue, no such path can exist, meaning that the outside of the iris is connected to $\mathscr A$ by a blue path.  We have recreated the final conditions after the switch and this case is proved.  All other cases are $\{$u$, $v$, $w$, \mbox{yellow, blue}\}$ permutations and discrete rotations of the two described above.  In starting with color neutral combinations we always end up (via a slightly different route than in the non--iris case) with color neutral combinations, and Cauchy--Riemann relations for these functions are established.
\qed    

\subsection{Contour Integration}
We now wish to show that the functions $u_{_N}$, $v_{_N}$ and $w_{_N}$ converge to limiting objects which are indeed harmonic.  We will do this by showing that the functions $u_{_N} - \tau^2 v_{_N}$, $v_{_N} - \tau^2 w_{_N}$ and $w_{_N} - \tau^2 u_{_N}$ converge to analytic functions via Morera's theorem.  Specifically, we first compute the contour integral around a single hexagon and show that this reduces to leftover derivative pieces.  These pieces are judiciously and symmetrically placed about the hexagon in such a way as to cancel leftovers from neighboring hexagons.  Hence, by discrete distortions, any contour integral around a region of $N^2$ hexagons will result in some derivative pieces around the contour which are easily shown to be small.  We start with some notation and a definition.

\begin{nota}\label{hex}
Hexagons are oriented as before, that is to say with two edges parallel to the $y$--axis.  We label the vertices of the hexagon counterclockwise starting with the bottom vertex by $z_1$, $z_2$, $z_3$, $z_4$, $z_5$, $z_6$.  If $f$ is a function defined on the lattice, then we may use the notation $f(z_i)$ or $f_i$ to denote the value of the function at the site $z_i$.
\end{nota}

\begin{defn}\label{cont int}
Let $\mathscrb C= \{z_1, \dots, z_n\}$ denote a contour consisting of neighboring points on the hexagonal lattice and $f$ a complex valued function on the hexagonal lattice.  Then we define the discrete contour integral via
\begin{equation*}
\oint^{N}_{\mathscrb C} f dz = \frac{1}{N} \sum_{k=1}^{n}  [f(z_k) + f(z_{k+1})]\cdot \frac{1}{2}\cdot(z_{k+1} - z_k).
\end{equation*}
That is to say, in our definition, the value of $f$ for the contour element is determined by \emph{both} endpoints of the bond.  Note that this has the advantage that integrations in the opposite directions of each contour element cancel exactly.  
\end{defn}

We remark that the factor of $\frac{1}{N}$ is for the anticipated scaling, so that the above display should be understood in the spirit of a contour whose length is of order $N$.  In the forthcoming lemma, we will deal with small scale contours so, to avoid introduction of additional notation, we transfer the $N$ to the other side of the equation:

\begin{lemma}\label{hex mess}
Let $\partial H$ denote the contour which is the boundary of a hexagon in accord with Notation \ref{hex}.  Then 
\begin{equation*}
N\oint_{\partial H}^N [u_{_N}^*(z) - \tau^2 v_{_N}^*(z)] dz = i (\alpha_{_H} + \tau \beta_{_H} + \tau^2 \gamma_{_H}),
\end{equation*} 
where $\alpha_H$, $\beta_H$ and $\gamma_H$ are real numbers that represent sums of derivative pieces of $u_{_N}^*$.  Furthermore, these functions have a tiling symmetry in the sense that e.g.~the quantity $\alpha_H$ associated with a particular hexagon $H$ is cancelled by the sum of the corresponding quantities $\alpha_{\tilde H}$ for all hexagons $\tilde H$ which neighbor the hexagon $H$; similarly for $\beta_H$ and $\gamma_H$.
\end{lemma}
\noindent
{\bf Proof: }We will provide a demonstration only for the case of the $\alpha_H$'s, since the situation for the $\beta$'s and $\gamma$'s are analogous.  An explicit calculation yields
\begin{equation*}\begin{split}
\alpha_{_H} &= [(u^*_2 - u^*_1) + (u^*_1 - u^*_6) + (u^*_3 - u^*_4) + (u^*_4 - u^*_5)]\\ &+ [(v^*_1 - v^*_6) + (v^*_6 - v^*_5) + (v^*_2 - v^*_3) + (v^*_3 - v^*_4)],\end{split}
\end{equation*}  
where, by the addition and subtraction of terms, the above has been written so that each term is a derivative along some edge of the hexagon.  Now we apply Lemma \ref{CR} and cancel off all corresponding pieces in such a way that everything is written in terms of the Cauchy--Riemann pieces of $u^*$.  We are then left with 
\begin{equation*}
\alpha_{_H} = [u^*_5]_{-\hat b}^+ + [u^*_5]_{-\hat c}^+ + [u^*_4]_{\hat c}^+ - [u^*_3]_{-\hat b}^+ - [u^*_2]_{\hat b}^+ -[u^*_2]_{\hat c}^+ - [u^*_1]_{-\hat c}^+ + [u^*_6]_{\hat b}^+. 
\end{equation*}
Associating, in a natural fashion, derivative pieces with the corresponding edge, it is seen that half of the corresponding edges are in $H$ and half of them ``invading'' a neighboring hexagon.  (So that in particular, there will be corresponding ``invasions'' from neighboring hexagons.)  It is not terribly difficult to see that each of the above pieces will occur in the integration of four hexagons, twice with positive sign and twice with negative sign and therefore cancel.
\qed

\begin{lemma}\label{cont int2}
Let $\Lambda^{(N)}_{\mathfrak F_N}$ denote a floral arrangement in a simply connected, regular region which has of order $N^2$ hexagons, and with boundary regions $\mathscr A$, $\mathscr B$ and $\mathscr C$, each of which is comprised of order $N$ hexagons.  Finally, let $\mathscrb C_{_N}$ denote a simple closed contour in $\Lambda^{(N)}_{\mathfrak F_N}$ whose length is also of order $N$.  Then there is some $\vartheta > 0$ and some constant $C_0 < \infty$, such that
\begin{equation*}
\left|\oint^N_{\mathscrb C_{_N}} [u^*_{_N}(z)-\tau^2v^*_{_N}(z)] dz\right| \leq C_0 N^{-\vartheta}, 
\end{equation*}
and similarly for $v^*_{_N} - \tau^2 w^*_{_N}$ and $w^*_{_N} - \tau^2 u^*_{_N}$.
\end{lemma}

\noindent
{\bf Proof: }We perform the contour integral in accord with the formula in Definition \ref{cont int} withholding the overall factor of $\frac{1}{N}$ for later purposes.  We may freely indent the contour one hexagon at a time, ultimately exhausting all interior hexagons.  Each interior hexagon, that is to say a hexagon which does not share at least one of its edges with $\mathscrb C_{_N}$, provides zero net contribution in accord with Lemma \ref{hex mess}.  What remain are the leftover Cauchy--Riemann pieces on or near the boundary, the number of terms of which is of order $|\mathscrb C_{_N}|$, which itself is of order $N$.  However, each piece corresponds to the probability of disjoint connections to the three boundary regions, at least one of which must be of order $N$ away.  Using the $4^{\mbox{\footnotesize{th}}}$ item in Theorem \ref{critical} the result follows.
\qed

\subsection{Proof of Theorem \ref{main theorem}}
For $\mathscrb Z \in \mathcal D$ let us denote by $U_N (\mathscrb Z)$ the function $u_{_N}(N \mathscrb Z)$, and similarly for $V_N(\mathscrb Z)$ and $W_N(\mathscrb Z)$.  While the statement of the theorem concerns the blue and yellow versions of these functions, here, for obvious reasons, we deploy the color--neutral objects.  In Corollary \ref{color neutral}, we will show that 
\[\lim_{N \rightarrow \infty} |u_{_N}^B (z) - u_{_N}^Y(z)| = 0,\]
for all $z$,  so that the various limiting objects may be identified.  As has been discussed, the discrete derivatives have been displayed as (differences of) probabilities of events which require connections between $\mathscrb Z$ and all three boundary components.  Thus, regardless of the particulars of the position of $\mathscrb Z$, the discrete derivative always requires at least one long arm emanating from (the lattice location of) $\mathscrb Z$.  By Theorem \ref{critical}, item four, this vanishes with an inverse power of $N$, which in terms implies a H\"older estimate which is uniform in $\mathscrb Z$ and $N$.  It follows that the $U$, $V$ and $W$ sequences are equicontinuous, and we can extract sub--sequential limits (along a mutual subsequence) which we denote by $U(\mathscrb Z)$, $V(\mathscrb Z)$ and $W(\mathscrb Z)$.  Letting $\mathscrb C \subset \mbox{int}(\mathcal D)$ denote any simple, closed curve which is rectifiable, we write  
\begin{equation*}
\oint_{\mathcal C} [U(\mathscrb Z)-\tau^2V(\mathscrb Z)] d\mathscrb{Z} = \lim_{N \rightarrow \infty} \oint_{\mathcal C_{_N}}^N [u_{_N}(z) - \tau^2v_{_N}(z)] dz,
\end{equation*}
and similarly for the $V$, $W$ and $W$, $U$ pairs.  
We wish to make use of Lemma \ref{cont int2}, but in order to do so we must replace $u$, $v$ and $w$ by their starred versions.  On the basis of Lemma \ref{all the same} in the Appendix, we find that $|u_{_N}(z) - u_{_N}^*(z)|$ tends to zero uniformly for any particular contour, and similarly for $v$ and $w$.  This allows us to bring Lemma \ref{cont int2} into play and we may now assert that the limiting contour integrals vanish.  

By Morera's Theorem, it is evident that $U$, $V$ and $W$ are an ``analytic triple'', i.e. the functions $U+ i\cdot\frac{1}{\sqrt{3}}(V - W)$, $V+ i\cdot\frac{1}{\sqrt{3}}(W - U)$ and $W+ i\cdot\frac{1}{\sqrt{3}}(U - V)$ are all analytic.  However, it is immediately clear that these functions are not independent.  Indeed, upon addition of the three, the imaginary part of these vanishes, allowing us to conclude that $U+V+W$ is a constant, which, momentarily, we will show is unity.  Thus there is actually only one analytic function in play, e.g.~$U+V + i \cdot \frac{1}{\sqrt{3}}(U-V)$.  However, we will still have occasion to exploit the symmetry of the triple.  

The boundary values are inherited from the discrete lattice versions of these functions: $U = 0$ on $\mathscr C$, $V=0$ on $\mathscr A$ and $W=0$ on $\mathscr B$;  furthermore, at the point $e_{_{AB}}$ which joins the $\mathscr A$ and $\mathscr B$ boundaries, $U=1$, and similarly for $V$ and $W$ at the other junctures.  These are readily proved by another appeal to Theorem \ref{critical}, item four.  For example, let us consider the function $U(\mathscrb Z)$, with the point $\mathscrb Z$ in the midst of $\mathscr C$.  Then back on the discrete level, for all intents and purposes, this point must be joined to some point on $\mathscr A$ and another on $\mathscr B$ by blue transmissions.  Since $\mathscrb Z$ cannot be close to both boundaries, this probability tends to zero as $N$ tends to infinity.  Moreover, this argument is not confined to points that are actually on the boundary, a similar argument also demonstrates that for points near the boundary -- on the macroscopic scale -- $u_{_N}(z)$ takes on a small value.  Similar arguments hold for the boundary values of $V$ and $W$ on $\mathscr A$ and $\mathscr B$, and it is also not hard to show that as $\mathscrb Z$ approaches $e_{_{AB}}$, $U(\mathscrb Z)$ must approach one.
 
We claim that the boundary condition (and the symmetry of the triple) is, in fact, enough to specify uniquely what the function is -- namely the conformal transformation of the linear Cardy--Carleson function described in the introduction.  To establish this, it is sufficient to demonstrate that a similar sort of analytic triple laden with the constraint of adding up to zero -- i.e.~homogenous boundary conditions -- is identically zero.  We proceed as follows: Since all functions described are harmonic, we may, by conformal invariance, treat the corresponding (homogeneous) problem on a triangle.  On the triangle we denote the three functions as $\delta U$, $\delta V$ and $\delta W$ and, without loss of generality, $\delta U = 0$ leg of the triangle coincides with the $x$--axis.  Noting that $\delta U$ is the imaginary part of an analytic function, $\Phi_U$, whose real part is $-\frac{1}{\sqrt{3}} (2\delta V + \delta U)$, we may use the Schwarz Reflection Principle to extend this analytic function across the $x$--axis.  We will use the continuation of $\Phi_U$ to define a $\delta U$ and $\delta V$ throughout the reflected domain, i.e.~Im$(\Phi_U) =_{df} \delta U$ and $\frac{1}{2}[-\sqrt{3}\mbox{Re}(\Phi_U) - \mbox{Im}(\Phi_U)] =_{df} \delta V$.  It is found, obviously, that $\delta U$ changes sign upon this reflection.  More significantly, $\delta V$ takes on the reflection of the value $\delta U+\delta V$ which by the homogeneity assumption is exactly $-\delta W$, so $\delta W$ is given by the negative of the reflection of $\delta V$.  The boundary conditions on the new, reflected boundaries are therefore conditions that the (extended) $\delta V$ and $\delta W$ vanish.  A similar phenomenon will happen when reflecting across the $\delta V=0$ lines and/or the $\delta W=0$ lines.  It is therefore clear that starting from a triangle whose indefinite reflections will tile the plane, e.g.~a right triangle or an equilateral triangle, we will end up with a triplet of analytic functions whose individual components are always, to within a sign, one of the original $U$, $V$ or $W$ evaluated at the corresponding point in the original triangle.  It is evident that these functions are all bounded and, often enough, zero, so they are all identically zero.  

Since the subsequence led to an unambiguous limit we conclude convergence of the full sequence, and the desired result has been established.
\qed

\section{Conclusion}
We have studied a model which differs in no outstanding way from any other in a myriad of 2D percolation models.  We demonstrated that, at least as far as the crossing probabilities are concerned, the continuum limit of the present model is identical to that of the site model on the triangular lattice.  Needless to say, there are obvious similarities between the present model and the site model on the triangular lattice -- in particular, vis--a--vis a hexagonal tiling problem.  (Not to mention that the model without irises is the $s=0$ limit of the model with irises.) All in all, these similarities allowed for the development of a proof which follows closely the original derivation of \cite{Smir}.  Notwithstanding, a small amount     -- but one which is of strictly positive measure -- of universality has been established.  In particular, and of similarly small significance, is the fact that the parameter $s$ may take on a range of values and needless to say, there is a good deal of leeway in the placement of flowers.    

There are numerous shortcomings to this work.  It is worthwhile to underscore the ones which we believe are of greater significance: \\
\indent 1.  It has not proven feasible for us to establish these results for well--known systems.  In particular, one has in mind, among the self--dual problems, the full bond triangular lattice and/or the acclaimed bond problem on the square lattice, not to mention any number of 2D critical models without self--duality.  We envision that in the former sorts of systems, an approach akin to the existing techniques might be developed, while for the latter, perhaps, an entirely new approached will be required.\\
\indent 2.  While the touted advantage of a derivation along the lines in \cite{Smir} is the demonstrated robustness of the approach, the downside is that the present work sheds no new light on the nature of the critical phenomena.  For example, while anticipated that the Cauchy--Riemann equations should become manifest on a mesoscopic scale, at least as far as the authors' current understanding goes, they appear to obscure with any deviation from the microscopic hexagonal geometry.\\
\indent 3.  On a more specific note, the authors find it highly regrettable that a rigid flower arrangement was required.  In point of fact, all of the essential results, e.g.~color parity of the transmission probabilities, Cauchy--Riemann relations, etc.~were established for entirely arbitrary flower arrangement.  What could not be done, at least not without additional labor, was the establishment of the standard critical properties of a 2D percolation system.  Here, it appears (after all these years) that some significant form of lattice symmetry is still required.  Notwithstanding, the authors envision a stochastic version of the current system.  For example, the presence or absence of an iris could be governed by a local random variable and the values of $s$ within the iris may also be random variables.  Under some reasonable homogeneity assumptions, such problems might be approached by methods along the lines of the present work. 

Finally (and one might presume that this is eminently rectifiable) would be the completion of the preliminary description for the continuum limit of this model by making the connection to SLE$_6$.  This topic is under consideration and may very well be the subject of a later paper.            

\section{\large{Appendix 1: Harris--FKG Properties and Criticality}}
Here we give a proof of the FKG property needed to prove Corollary \ref{critical}.  We point out that in the strict sense our model does not enjoy positive correlations, as the following example shows:
\begin{ex}
Consider a single flower with the petals labeled as in Section \ref{def}.  Let $S_{\{4, 5\}}$ be the set containing petals 4 and 5 and let $S_{\{1\}}$ denote the singleton set containing petal 1.  Let $\{S_{\{4, 5\}} \leftrightarrow S_{\{1\}}\}$ denote the event of a connection between $S_{\{4, 5\}}$ and $S_{\{1\}}$.  Then it is claimed:
\begin{equation}\label{ineq}
\mathbb P(\{S_{\{4, 5\}} \leftrightarrow S_{\{1\}}\} \mid S_{\{4, 5\}} =  S_{\{1\}} = B) < \mathbb P(\{S_{\{4, 5\}} \leftrightarrow S_{\{1\}}\}).
\end{equation}
Let us start by conditioning on the state of petal 6.  The conditional probability given that petal 6 is blue is 1 for both the left hand side and the right hand side of Eq.~(\ref{ineq}), so we might as well consider the case where petal 6 is yellow.  Let us start with the unconditioned probability, i.e.~the right hand side.  It is claimed that, as far as the rest of the petals are concerned, there are three scenarios: predetermined transmission (i.e.~a connection without use of the iris), a trigger and other.  The relevant conditional probabilities are 1, $\frac{1}{2}$ and $a+2s$, respectively, with the exception of a single configuration which is in both categories (i) and (ii).  The resultant tally is: 
\begin{equation}\label{yallow1}
\mathbb P(\{S_{\{4, 5\}} \leftrightarrow S_{\{1\}}\} \mid S_{\{6\}} = Y) = 2^{-5} \left[5 \cdot \frac{1}{2} + 8 + 19 \cdot (a + 2s)\right].
\end{equation}
For the conditional probability, we simply calculate all four cases, with the result:
\begin{equation}\label{yallow2}
 \mathbb P(\{S_{\{4, 5\}} \leftrightarrow S_{\{1\}}\} \mid \{S_{\{4, 5\}} =  S_{\{1\}} = B\} \cap \{S_{\{6\}} = Y\}) = \frac{1}{4} \left[1 + \frac{1}{2} + 2(a+2s)\right]. 
\end{equation}
By repeated use of the fact that $2a+3s = 1$, it is seen that the right hand side of Eq.~(\ref{yallow1}) exceeds the right hand side of Eq.~(\ref{yallow2}) whenever $s>0$.
\end{ex}

However, for the purposes of proving criticality we in fact only need positive correlations on paths.  More precisely, we have
\begin{lemma}\label{FKG}
Let $\Lambda_{\mathscr F}$ denote a flower arrangement and let $A_1, B_1; A_2, B_2; \dots A_n, B_n$ denote sets in $\Lambda_{\mathscr F}$ in the complement of irises.  Let $\mathbb T_1$ denote the event that $A_1$ and $B_1$ are blue and that $A_1$ is connected to $B_1$ by a blue path, with similar definitions for $\mathbb T_2, \dots, \mathbb T_n$.  Then, under the condition that $a^2 \geq 2s^2$, the events $\mathbb T_1, \dots, \mathbb T_n$ are all positively correlated, i.e., if $J \subset \{1, 2, \dots, n\}$ and $L \subset \{1, 2, \dots, n\}$ then 
\begin{equation*}
\mu_{\Lambda_{\mathscr F}}(\bigcap_{j \in J} \mathbb T_j \cap \bigcap_{\ell \in L} \mathbb T_\ell) \geq \mu_{\Lambda_{\mathscr F}} (\bigcap_{j \in J} \mathbb T_j) \mu_{\Lambda_{\mathscr F}} (\bigcap_{\ell \in L} \mathbb T_\ell)
\end{equation*}
\end{lemma}

\noindent
{\bf Proof: }We consider first the binary case -- multiple path cases following an nearly identical argument.  Let $\sigma$ denote a generic configuration of petals and filler and let $I$ denote a generic  configuration of irises.  Our first claim is that the function 
\begin{equation*}
T_j(\sigma) = \mathbb P_{\Lambda_{\mathscr F}} (\mathbb T_j \mid \sigma)
\end{equation*}
is an increasing function of $\sigma$.  To see this, let $\sigma$ and $\sigma \vee \eta$ denote configurations which differ only at the site $\eta$ -- where the latter is blue and the former is yellow.  If $\eta$ is a filler site the claim is obvious.  Similarly, if $\eta$ is a petal site where the presence/absence of blue does not affect the trigger status of the flower, the result is also trivial.  Futhermore, it is also clear that if the path event does not depend on the iris (i.e.~if the iris is not a pivotal site for the event $\mathbb T_j$) then the raise at $\eta$ can no deleterious effect on $\mathbb T_j$.  Thus we must only consider situations where the state of $\eta$ causes or disrupts a trigger and a transmission through the iris is crucial for the event that $\mathbb T_j$ occurs.  

First we consider the case where raising at $\eta$ leads to a triggering situations.  In this case, the associated flower must have started with exactly two blue petals.  If the two blue petals were already adjacent then it is obvious that the raise at $\eta$ can only benefit the possibility of the event $\mathbb T_j$, i.e., assuming the cooperation of the iris, this could complete a connection.  Let us consider the case where the blue petals were not adjacent.  We must resort to considering the full event $\mathbb T_j$ on the configuration $\omega = (\sigma, I)$.  We must thus compare the (conditional) probability of a connection between our blue petals of $\sigma$ (without the trigger) and our three blue petals of $\sigma \vee \eta$ with the trigger.  The latter is $\frac{1}{2}$ while the former is $a+s < \frac{1}{2}$.  Now we turn to the case where the raise at $\eta$ disrupts a trigger.  Before the raise, the connection probability is $\frac{1}{2}$ whereas after the raise, the connection probability is either 1 (because the two relevant sets get connected outside the iris) or, in the two less trivial cases, $a + 2s > \frac{1}{2}$.  So our first claim is established.

We note that the conditional measure $\mu_{\Lambda_{\mathscr F}} (- \mid \sigma)$ (for whom the only degrees of freedom are represented by the iris configurations) is in fact independent -- but not necessarily identically distributed -- measure on the irises.  In \cite{CL} it was proved that in an analogous circumstance with parameters $a_i, e_i, s_i$, $i = 1, 2, \dots$, that provided $a_ie_i \geq 2s_i^2$ is satisfied for all $i$, the corresponding product measure has positive correlations.  This is our situation where some $a_i = e_i = \frac{1}{2}$ and otherwise $a_i e_i= a^2 \geq 2s^2 = 2s_i^2$ by hypothesis.  Since the indicator function of the event $\mathbb T_j$ is manifestly increasing in the iris configurations, we have correlation inequalities for the conditional measure; so
\begin{equation*}
\mathbb E(\mathbb T_j \mathbb T_l \mid \sigma) \geq \mathbb E(\mathbb T_j \mid \sigma) \mathbb E(\mathbb T_l \mid \sigma) = T_j(\sigma) T_l(\sigma).
\end{equation*}
The desired result follows by taking the expectation over petal/filler configurations and using the Harris--FKG property for independent percolation.  The proof for multiple path events as well as a variety of other increasing events follows mutatis mutantis from the argument given. 
\qed

\begin{remark}
With additional labor, it may be possible to remove the $a^2 \geq 2s^2$ restriction.  However, we shall not pursue this avenue since, in any case, we require that $a \geq \frac{1}{5}$.
\end{remark}

\section{\large{Appendix 2: Equivalence of the Cardy--Carleson Functions}}
In this appendix, we will supply the necessary details to show that the difference between our functions $u_{_N}^*(z)$, $v_{_N}^*(z)$ and $w_{_N}^*(z)$ are, for all intents and purposes, equal to the unstarred counterparts.  We start with some notation:

\begin{defn}
Let $B_n$ denote the $2n\times2n$ box centered at the origin -- that is to say all those hexagons within an $L^1$ distance $n$ of the origin -- and $\partial B_n$ the hexagons of $B_n^c$ with a neighbor in $B_n$.  While technically we should also specify the location of the origin relative to the flower arrangement, in what is to follow such amendments would only result in the adjustment of a few constants in some of the estimates.  We will not pay heed to these matters in the forthcoming definitions and the various later estimates should be understood as the maximum or minimum over a single period of translations.

Let $\Pi_1(n)$ denote the event that the origin is connected to $\partial B_n$  by a blue transmission and let $\pi_1(n)$ denote the corresponding probability.   Similarly, we consider multiple disjoint paths of various colors and arrangements which connect the origin to $\partial B_n$ and we use the subscript to indicate the number of paths with the color and arrangement dependence notationally suppressed.  Of importance will be the five--arm event, $\Pi_5(n)$, the subject of some discussion in \cite{AizChi},\cite{KSZ} and \cite{LSW} wherein the origin is connected to $\partial B_n$ by three blue paths and two yellow paths, with the two yellow paths separated by blues.  (In \cite{KSZ}, it was proved that the corresponding probability, $\pi_5(n)$, has upper and lower bounds of a constant divided by $n^2$; these arguments, at least the upper bounds, are easily adapted to the present circumstances.)  Next, if $m < n$, we define $\Pi_1(n,m)$ to be the event of a connection between $\partial B_m$ and $\partial B_n$ and we denote the corresponding probability by $\pi_1(n,m)$.  We adapt similar notations for $\pi$-functions involving multiple disjoint connections in the annular region.  Finally, we will consider versions of these events with a geometric restriction.  Let $\theta \in [0,2\pi)$ and consider the ray starting from the origin that makes angle $\theta$ with the horizontal axis. We define $\Pi_{1}^{\Bbb K,\theta}(n)$,  $\Pi_{2}^{\Bbb K,\theta}(n)$, $\dots$ to be the event that the appropriate paths occur subject to the constraint that none of the paths intersect the ray at angle $\theta$.  We use the same notation with a lower case $\pi$ to denote the relevant probabilities.  Similarly, we define $\Pi_{1}^{\Bbb K,\theta}(n,m)$, $\dots$ and $\pi_{1}^{\Bbb K,\theta}(n,m)$, \dots to denote the modified versions of the above mentioned for the annular regions $B_n\setminus B_m$.

We will also bring into play certain events of the type described in the above paragraph that incorporate additional events defined from the space of permissions.  These objects will be introduced as necessary.    
\end{defn}  
 
We begin with the central lemma of this appendix.  The proof relies heavily on asymptotic estimates of certain $\pi$-functions which will be proved in subsequent lemmas.
\begin{lemma}\label{all the same}
Let $u_{_N}^*$, $u_{_N}$ denote the functions as described previously, with domain $\Lambda$.  Let $\mathscrb Z$ denote a point in the interior of $\Lambda$, $z = N\mathscrb Z$.  Then, 
\begin{equation*}
\lim_{n \rightarrow \infty} |u_{_N}^* (z) - u_{_N}(z)| = 0.
\end{equation*}
In particular, on closed subsets of $\Lambda$ that are disjoint from the boundary, the above is uniformly bounded by a constant times an inverse power of $N$.
\end{lemma}
\noindent
{\bf Proof: }We claim (c.f.~below) that in those configurations in which $\mathscrb U_{_N}$ and $\mathscrb U_{_N}^*$ differ, a rather drastic event must occur involving multiple arms connected to the boundary and encircling $z$.  If this event occurs far away from $z$ and the boundary, then there are many, namely greater than five, long arms emanating from a single point.  By the modification of some above mentioned standard results, we can show that the instances of this event in the bulk, i.e.~away from the boundary and away from $z$, are suppressed.  On the other hand, when the path ventures near $z$ itself, not all of these arms will be long and, conditionally speaking, such a multi--arm event is not particularly unlikely.  However, the latter cases we claim are themselves unlikely; indeed most of the configurations contributing to $u_{_N}$ or $u_{_N}^*$ stay well away from $z$ on the microscopic scale.  Finally, for points near the boundary, while there may be fewer long arms to work with, the geometric constraints prove to be sufficient for our purposes.  The details are as follows:

Let us first consider the event which is contained in both the starred and unstarred versions of the u--functions, namely the event of a self--avoiding, non--self--touching path separating $z$ from $\mathscr C$, etc.  We will denote the indicator function of this event by $\mathscrb U_{_N}^-$.  Similarly, let us define an event, whose indicator is $\mathscrb U_{_N}^{*+}$, that contains both the starred and unstarred versions: this is the event that a separating path of the required type exists, with no restrictions on self--touching, and is allowed to share hexagons provided that permissions are granted.  It is obvious that 
\begin{equation}\label{tuff}
\mathbb E[\mathscrb U_{_N}^{*+} - \mathscrb U_{_N}^-] \geq |u_{_N}^*  - u_{_N}|.
\end{equation}

We turn to a description of the configurations, technically on $(\omega, X)$, for which $\mathscrb U_{_N}^{*+} = 1$ while $\mathscrb U_{_N}^- = 0$.  In such a configuration, the only separating paths contain an \emph{essential} lasso point which, we remind the reader, could be either a shared hexagon or a closed encounter pair.  For standing notation, we denote this ``point'' by $z_0$.  A variety of paths converge at $z_0$: certainly there is a blue path from $\mathscr A$, a blue path to $\mathscr B$, and an additional loop starting from $z_0$ (or its immediate vicinity) which contains $z$ in its interior.  However, since the lasso point was deemed to be essential, there can be neither a blue connection between this loop and the portion of the path connecting $z_0$ to $\mathscr A$ nor a blue connection between this loop and the portion of the path connecting $z_0$ to $\mathscr B$.  This implies two additional yellow arms emanating from the immediate vicinity of $z_0$.  These yellow arms may themselves encircle the blue loop and/or terminate at either  the two boundaries $\mathscr A$ and $\mathscr B$.  We remark that, specifying the lasso point under study to be the first (and by the same token the final) such point on the blue journey from $\mathscr A$ to $\mathscr B$, the paths from the boundaries to $z_0$ as well as the yellow paths mentioned have no sharing and, without loss of generality, no points of close encounter.  While such claims cannot be made about the loop, it is already clear that there are ``somewhat more'' than five standard arms emanating from the vicinity of $z_0$.  Turning attention to this blue loop, let us regard this as two separate paths -- with possible sharings -- each portion of which visits all the essential lasso points; the break between the two paths may be chosen arbitrarily after the final lasso point just prior to the capture of $z$.  Now we may claim that on the basis of Lemma \ref{loop erasure}, one of these two paths may be reduced to a self--avoiding and non--self--touching path.  Thus, to summarize, there are in fact six paths emanating from $z_0$; a pair of blue paths separated from another pair of blue paths by a pair of yellow paths.  One of the blue pairs is completely ``normal''.  The other blue pair, ostensibly two halves of a loop, will be regarded as one normal path and a second path which has received permissions to share and/or experience close encounters with the first.

Notwithstanding, the blue pair which captures $z$ along with a surrounding yellow loop cannot \emph{a priori} be ruled as unlikely if $z_0$ is in the vicinity of $z$.  To handle such points we let $0<\lambda < 1$ denote a number to be specified momentarily.  We now define $z_0$ to be  ``near''  $z$ if it is within a box of side $N^\lambda$ centered at $z$.  Since $\mathscrb Z \in$ int$(\Lambda)$, $z$ itself is a distance of order $N$ from the boundary.  Such an event would thus require a connection between the boundary of the above mentioned box to the outside of a larger box, also centered at $z$, which is the smallest such box that will fit in $\Lambda$.  This, for $N$ large enough, is a translation of the event $\Pi_1(d_{\mathscrb{Z}}N,N^\lambda)$ where $d_{\mathscrb{Z}}$ is a constant related to the distance between $\mathscrb{Z}$ and the boundary of the domain measured on the unit scale.  By standard arguments employing rings in disjoint annuli (which go back to \cite{Harris}) we may, on the basis of Theorem \ref{critical}, show that the probability of such an event is bounded above by a constant times $(\frac{N^\lambda}{N})^{\vartheta_1}$ for some $\vartheta_1 > 0$.

Hence for all intents and purposes, when we examine the configurations where $\mathscrb U_{_N}$ and $\mathscrb U_{_N}^*$ are purported to differ, we may assume that there is no visit to the near vicinity of $z$.  (In particular, we certainly need not worry about the fractional values of $\mathscrb U_{_N}^*(z)$ when the path goes directly through $z$.)  Furthermore we will now regard, 
with only small loss of generality, the expectation in Eq.~(\ref{tuff}) to be taking place in the conditional measure where no path from the boundary visits the near vicinity of $z$.  It follows that for $z_0$ located anywhere in $\Lambda$ a distance further than $N^\lambda$ from the boundary (and $z$) all of the above mentioned paths emanating from the vicinity of $z_0$ travel to the outside of a box of side $N^\lambda$ centered at $z_0$.  We denote the probability of this modified six--arm event by $\pi_{6^*}(N^{\lambda})$.

In light of \cite{KSZ}, it should come as no surprise that
\begin{equation}
\label{six-star}
\pi_{6^*}(N^{\lambda}) \leq  
\frac{C_{6^*}}{N^{\lambda(2 + \vartheta_2)}}
\end{equation}
with $C_{6^*}$ a number of order unity independent of $N$ and $\vartheta_2 >0$.
In any case, the inequality in Eq.(\ref{six-star}) is the subject of Lemma \ref{six-star lemma}. Thus, choosing $\lambda$ close enough to one to ensure that the power in the denominator of the right hand side exceeds two, we may sum over all relevant values of $z_0$ and thereby dispense with the so-called bulk terms.

This leaves us with the boundary contribution which we divide into two (technically three) types.  First there are points which lie near a corner of the domain and then there is the complementary set.  Along with the former, we will include the points near the juncture of the $\mathscr A$--$\mathscr B$ boundary i.e.~the point $e_{_{AB}}$.  Since there are only a finite number of these sorts of boundary points and the associated nearby points are handled rather easily, let us define our ``vicinity'' of these points and dispose of these regions immediately.  

We let $\mu_2$ be a number larger than $\lambda$ but still smaller than one: $1 > \mu_2 > \lambda$, and at each corner, we place a box of side $N^{\mu_2}$ (with its center at the corner) and another such box at  $e_{_{AB}}$.  If $z_0$ lies inside one of these boxes, some of the six arms will still be long.  In particular, for future reference, concerning the corner points of the $\mathscr A$ boundary or the $\mathscr B$ boundary that are distinct from $e_{_{AB}}$, there are at least four long arms.  As it turns out, the points near $e_{_{AB}}$ have two.  Regardless of the exact tally, it is clear that, for each such point mentioned, if $z_0$ is in the associated box, the boundary of this box must be connected a distance of order $N$ and so 
the requisite event is contained in a translate of the event $\Pi_1(k N, N^{\mu_2})$.  Here $k$ is some constant of order unity independent of $N$ which can again be related to various distances in unit scale domain.  Hence we pick up a finite number of additional terms with the upper bound of a constant times $(\frac{N^{\mu_2}}{N})^{\vartheta_1}$. 

Finally there is the remainder of the points near the boundary:  points that are within a distance $N^{\lambda}$ of the boundary but further than $N^{\mu_{2}}$ from any of the corners or $e_{_{AB}}$.  By definition, if we place a box of side exceeding $2N^\lambda$ of any of these points, that box will intersect $\Lambda^c$.  Thus let us cover this region with partially overlapping boxes of side, say, $3N^\lambda$ and notice that the number of boxes is of the order $N^{1-\lambda}$.  Further, it is noted that, on a distance scale of $N^\lambda$, all these boxes are well away from all the corners.  Thus the boundary region near any particular box is, essentially, a straight edge and there is ample room to draw straight lines in the complement of $\Lambda$ which start from the boundary of these boxes, are directed towards their centers, and are large compared with $N^\lambda$ but, perhaps, small compared with $N$.  

We now take each of the above mentioned boxes and place it at the center of a box of side $2N^{\mu_1}$, where $\mu_2 > \mu_1 > \lambda$.  As indicated above, we can connect the boundaries of these boxes by a straight line which lies in $\Lambda^c$ and is directed towards their mutual center.  If $z_0$ is inside the inner box, then, as alluded to earlier, there must be four arms which connect the boundary of the inner box to the boundary of the outer box.  Two of these four arms are yellow and two of these are blue, with the pair of blue arms between the yellow arms; the yellows and one of the blues are self--avoiding and non--self--touching while the second blue interacts with the first given the requisite permissions.  In short, four of the six arms that were dealt with in the context of the bulk contribution.  However, clearly these arms are restricted so as not to enter the region $\Lambda^c$; certainly they cannot cross the straight line described in the above paragraph.  The relevant event is therefore $\Pi_{4^*}^{\mathbb K, \theta}(N^{\mu_1}, \frac{3}{2}N^\lambda)$, where $4^*$ means pretty much what $6^*$ meant in the earlier context.

The subject of Lemma \ref{ice pick lemma} is that for the usual three arm version of the above event, $\pi_{3}^{\mathbb K, \theta}(n, m)$, has upper and lower bounds of the form a constant times $m/n$, where the constant is uniform in $\theta$.  Therefore it once again should not be surprising that 
\begin{equation}\label{ice}
\pi_{4^*}^{\mathbb K, \theta}\left(N^{\mu_1}, \frac{3}{2}N^\lambda\right) \leq C_{4^*} \left(\frac{N^\lambda}{N^{\mu_1}}\right)^{1+\vartheta_3},
\end{equation}
with $\vartheta_3 > 0$ and $C_{4^*}$ a constant.  The estimate in Eq. (\ref{ice}) will be proved as a corollary to Lemma \ref{ice pick lemma}.   

Summing over all such boxes, the overall remaining contribution is therefore no more than a constant times $N^{1 + \lambda \vartheta_3 - \mu_1(1 + \vartheta_3)}$.  The above exponent is negative if we choose (first $\mu_2$ and then) $\mu_1$ sufficiently close to one.  It is not difficult to ascertain that every one of the above estimates are uniform in $z$ provided that $z$ remains a fixed distance from the boundary.  The lemma is proved.
\qed    

\begin{lemma}\label{six-star lemma}
Consider the event $\Pi_{6^*}(n)$ as described in the proof of Lemma \ref{all the same} and let $\pi_{6^*}(n)$ denote the corresponding probability.  Then, for all $n$, there is a finite constant $C_{6^*}$ which does not depend on $n$, such that
\[\pi_{6^*}(n) \leq  
\frac{C_{6^*}}{n^{2 + \vartheta_2}}.\]
\end{lemma}    
\noindent
{\bf Proof: }We start with some discussions concerning the five--arm event $\Pi_5(n)$, which, in the present circumstances, means two yellow paths and three blue paths with the two yellow paths separated.  According to the arguments of Lemma 5 in \cite{KSZ}, the probability of a \emph{particular arrangement} of the five arms (certain arms ending up at certain boundaries, etc.) is easily bounded above by a constant times $n^{-2}$.  This argument goes through intact for the systems under consideration in this work.  The crux of the matter is, therefore, to show that with conditional probability of order unity the system will end up in the preferred arrangement.  This rather difficult matter was first resolved for the four--arm case in \cite{Kesten87} and indeed this resolution was the technical core of that work.  Most of the intricate construction consisting of fences, corridors, etc.~relies on standard critical properties of 2D percolation models, specifically the second and third items in Theorem \ref{critical}.  We remark that there were numerous points in the derivation where the Harris--FKG inequalities were employed.  In essentially all of these cases, Lemma \ref{FKG} applies directly, as the relevant events always involved paths and connections.  A small exception consists of Lemma 3.  Here the proof in \cite{Kesten87} would go through intact provided that the disjoint regions in question were in fact ``flower disjoint'', e.g.~in the notation of \cite{Kesten87}, the sets ``$\mathscr A$'' and ``$\mathscr E$'' must contain no flower in common.  These and similar conditions for related sets can be arranged in any number of ways; to be specific, in every square and rectangle on all of the various scales, one may ``waste'' a buffer zone layer whose thickness consist of at least one unit cell.  Needless to say, certain modifications of the four--arm argument must be made for the benefit of five and further arms -- here the issue being that in the five arm cases, the colors no longer alternate.  These matters were discussed in Section 7 (Appendix to Lemma 5) of \cite{KSZ}.  The arguments therein can be applied with almost no modification.

To prove Eq.~(\ref{six-star}) one should, ostensibly, employ some sort of disjoint occurrence argument.  Unfortunately the modern versions, e.g.~Reimer's inequality, do not appear to be readily adapted to the current set up, so we must resort to old fashioned methods of conditioning.  We claim that in fact $\pi_{6^*}(n) \leq \pi_5(n) \pi_1(n)$.  Let us label the yellow arms $Y_1$ and $Y_2$, as ordered counterclockwise, with the ``loop arms'' between them.  Calling the ``normal'' arm of the loop $B_1$ we envision the second loop arm as lying between $B_1$ and $Y_2$.  We now condition on the clockwise--most transmission for the arm $B_1$ and counterclockwise--most transmission for the arm $Y_2$.  We denote the region in between by $\mathscr R_{B_1, Y_2}$ and, with apologies, the extreme versions of these paths by $B_1$ and $Y_2$, respectively.    

Were it not for the possibility of sharing, our conclusion is immediate.  We underscore that there are two forms of sharing involved:~the mixed hexagons in $Y_1$ and the sharings with permission in $B_1$.  However, in the former case (c.f.~the proof of Lemma \ref{CR} for non--iris sites), and certainly in the latter case, we need not reveal which hexagons are available for sharing in order to provide the conditioning.  The content of Lemma \ref{full flower better} is that any path event, blue or yellow, has a greater probability in an unused flower than in a flower which has some parts conditioned on, notwithstanding that its iris may be available for sharing.  It is therefore manifest that in the region $\mathscr R_{B_1, Y_2}$ expanded by all the flowers of $B_1$ and $Y_2$, the probability of an additional blue transmission is, in fact, greater than the requisite transmission which actually has to receive permission (and does not get rejected for illicit close encounters).  However the probability in the above stated region is obviously less than $\pi_1(n)$; summing over all partitions -- and using the standard power law bounds on $\pi_1(n)$ -- provides us with the desired result.    
\qed        

\begin{cor}\label{color neutral}
Let $u_{_N}^B(z)$ and $u_{_N}^Y(z)$ denote the blue and yellow components of the function $u_{_N}$.  Then for all $z \in \mathcal D$, 
\[ \lim_{N \rightarrow \infty} |u_{_N}^B(z) - u_{_N}^Y(z)| = 0,
\]
with similar results for $v$ and $w$.
\end{cor}
\noindent
{\bf Proof: }While ostensibly it would seem that under the auspices of Lemma \ref{unconditioned}, the equality of $u_{_N}^B(z)$ and $u_{_N}^Y(z)$ is a forgone conclusion, it is conceivable that a difference might arise due to the disparity between the geometry of a path designate and the geometry of the transmission which achieves this designation.  However, the conditions under which this disparity might emerge are akin to the conditions which were shown to be vanishingly small in Lemma \ref{six-star lemma}.  In particular, this might happen if the designate goes directly through $z$ -- which happens to be in a flower, or, more pertinently, the path designate may contain a long loop capturing $z$ which is achieved by a realization making no use of this essential loop.  However, if this is to happen \emph{and} the underlying realization does not achieve the event $\mathscrb U_N(z)$, then we are back to a $\Pi_{6*}$--type event.      

To be specific, let $\mathscr T_{u_{_N}^B(z)}$ denote the collection of path designates which may be realized by a path from $\mathscr A$ to $\mathscr B$ separating $z$ from $\mathscr C$.  By our usual abuse of notation, we also use $\mathscr T_{u_{_N}^B(z)}$ to denote the event that some designate in this set is achieved by a blue transmission.  We define a similar quantity for yellow and, as a consequence of the arguments which were used in the proof of Lemma \ref{unconditioned}, 
\[ \mathbb P(\mathscr T_{u_{_N}^B(z)}) = \mathbb P(\mathscr T_{u_{_N}^Y(z)}).\]
On the one hand, it is clear that 
\[u_{_N}^B(z) \leq \mathbb P(\mathscr T_{u_{_N}^B(z)}).\]
Now let $\Xi_N(z)$ denote the complement of the events that were treated in Lemma \ref{six-star lemma}; e.g.~no blue path from the boundary visits the near vicinity of $z$, no $\Pi_{6*}$--type events, etc.  Then, on the other hand, from the above discussion, it is not difficult to see that
\[ u_{_N}^B(z) \geq \mathbb P(\mathscr T_{u_{_N}^B(z)} \mid \Xi_N(z)).\] 
The preceding pair of inequalities also hold with $B$ replaced by $Y$.  On the basis of the arguments used in the proof of Lemma \ref{six-star lemma}, we have $\mathbb P(\Xi_N(z)) \rightarrow 1$ as $N \rightarrow \infty$ and the desired result follows. 
\qed

\begin{lemma}\label{ice pick lemma}
Consider the events $\Pi_3^{\mathbb K, \theta}(n,m)$ as described in the proof of Lemma \ref{all the same} with $\pi_3^{\mathbb K, \theta}(n,m)$ the corresponding probability.  Then
\begin{equation*}
 C_3^\prime \frac{m}{n} \leq \pi_{3}^{\mathbb K, \theta}(n, m) \leq C_3 \frac{m}{n},
\end{equation*}
where $C_3$ and $C_3^\prime$ are constants independent of all parameters, including $\theta$.   
\end{lemma}

\begin{remark}
While the proof below is tailored to the system at hand, these ideas can obviously be generalized to a variety of critical 2D percolation models.
\end{remark}

\noindent
{\bf Proof: }We first assert that for fixed $r \in (0, 1)$, as $n \rightarrow \infty$, there exists a $\phi(r)$ such that
\begin{equation}\label{big small}
\pi_3^{\mathbb K, \theta}(n) \leq \phi(r) \pi_3^{\mathbb K, \theta}(rn),
\end{equation}
where the argument of the $\pi$ on the right--hand side is understood to mean a convenient integer value.  This can be established by making use of Kesten's fences (\cite{Kesten87}); however with only three arms it is not terribly difficult to construct an argument directly.  

Now consider the box $B_n$ with a line segment at angle $\theta$ cutting through the center of the box.  Let us assume for simplicity that the segment touches only two boundaries; one of these boundaries we will denote by $\mathfrak c$ and the rest of the boundary will be split into two parts by the ray, and we denote these parts by $\mathfrak a$ and $\mathfrak b$.  We parametrize the line segment by $\lambda$, where $\lambda = 0$ corresponds to the joining of the $\mathfrak a$ and $\mathfrak b$ boundaries and $\lambda = 1$ corresponds to the $\mathfrak c$ boundary.  Furthermore, we discretize the parametrization: $\lambda \in (\lambda_1, \dots, \lambda_k)$ so that the portion of the line segment corresponding to $\lambda_{j+1}$ contains one more hexagon than the the portion corresponding to $\lambda_j$.  We now define the event 
\[\begin{split}
\mathbb F(\lambda) = \{\omega \mid \exists 
&\mbox{ blue transmit from $\mathfrak a$ to $\mathfrak b$ which does}\\ 
&\mbox{ not cross the portion of the line segment}\\
&\mbox{ corresponding to parameter values in $[0, \lambda]$}\},
\end{split}\]
and we further define
\[f(\lambda) = \mathbb P(\mathbb F(\lambda)).\]

It is obvious that $f$ is monotone non--increasing in $\lambda$.  In fact, it is readily established that $f$ is \emph{strictly} decreasing since if $1>\lambda^\prime >\lambda>0$, it is possible, using corridors, to produce configurations of uniformly positive probability for which the $\mathbb F(\lambda)$ occurs while the event $\mathbb F(\lambda^\prime)$ does not.  We next observe that any $\omega \in \mathbb F(\lambda_{j-1}) \setminus \mathbb F(\lambda_j)$ for all intents and purposes lies in the restricted three--arm event in question.  In particular, in light of Eq.~(\ref{big small}) and another relocation of arms argument, for $\lambda_j$ not too close to zero or one, 
\[L_3 \pi_3^{\mathbb K, \theta}(n) \leq f(\lambda_{j-1}) - f(\lambda_j) \leq K_3 \pi_3^{\mathbb K, \theta}(n), \]
where $K_3$ and $L_3$ maybe regarded as independent of $\lambda$ for, say,  $\lambda \in (\frac{1}{4}, \frac{3}{4})$.  Summing up over the values of $\lambda$ in the above specified range, we learn that $\pi_3^{\mathbb K, \theta}(n)$ has upper and lower bounds of a constant times $n^{-1}$.  

To obtain the full stated result, we note that, clearly, 
\[\pi_3^{\mathbb K, \theta}(n) \leq \pi_3^{\mathbb K, \theta}(m) \cdot \pi_3^{\mathbb K, \theta}(n, m).\] 
However, invoking the techniques of \cite{Kesten87}, this may be supplement with a bound of the opposite type augmented by constants, which establishes the desired result.
\qed

\begin{cor}
Consider the function $\pi_{4^*}^{\mathbb K, \theta}(n, m)$ as described in the proof of Lemma \ref{all the same}, then 
\[\pi_{4^*}^{\mathbb K, \theta}(n,m) \leq c_{4^*}\left(\frac{m}{n}\right)^{1+\vartheta_3},\]
for some $\vartheta_3 > 0$.
\end{cor}
\noindent
{\bf Proof: }We use the result of Lemma \ref{ice pick lemma} in conjunction with a conditioning argument of the sort used in the proof of Lemma \ref{six-star lemma} to obtain this result.
\qed

\section*{\large{Acknowledgments}} 
We would like to acknowledge useful conversations with Jonathan Handy, Marek Biskup, John Garnett, and Christoph Thiele concerning the uniqueness of the functions $h_{\mathscr A}$, $h_{\mathscr B}$ and $h_{\mathscr C}$ on the basis of the existing boundary conditions.
\bigskip

\noindent
This work was in part supported by NSF under the grant DMS-0306167.

\end{document}